\documentclass[preprints,review,accept,moreauthors,pdftex]{Definitions/mdpi}

\firstpage{1} 
\pubvolume{7}
\issuenum{9}
\articlenumber{345}
\pubyear{2021}
\copyrightyear{2021}
\externaleditor{Academic Editor: Ezio Caroli }
\datereceived{18 June 2021} 
\dateaccepted{31 August 2021} 
\datepublished{14 September 2021} 
\hreflink{} 
\pdfoutput=1

\usepackage{acronym}
\usepackage{bm}
\usepackage{lscape}
\usepackage[normalem]{ulem}
\usepackage{bigints}

\Title{Probing Quantum Gravity with Imaging Atmospheric Cherenkov Telescopes}

\TitleCitation{Probing Quantum Gravity with Imaging Atmospheric Cherenkov Telescopes}

\Author{Tomislav Terzi\'{c} $^{1,}$*\orcidA{}, Daniel Kerszberg $^{2,}$*\orcidB{} and Jelena Stri\v{s}kovi\'{c} $^{3,}$*\orcidC{}}

\AuthorNames{Tomislav Terzi\'{c}, Daniel Kerszberg, Jelena Stri\v{s}kovi\'{c}}

\AuthorCitation{Terzi\'{c}, T.; Kerszberg, D.; Stri\v{s}kovi\'{c}, J.}

\address{%
$^{1}$ \quad University of Rijeka, Department of Physics, 51000 Rijeka, Croatia\\
$^{2}$ \quad Institut de F\'{i}sica d'Altes Energies (IFAE), The Barcelona Institute of Science and Technology (BIST),\linebreak E-08193 Bellaterra (Barcelona), Spain\\
$^{3}$ \quad Josip Juraj Strossmayer University of Osijek, Department of Physics, 31000 Osijek, Croatia}

\corres{Correspondence: tterzic@phy.uniri.hr (T.T.); dkerszberg@ifae.es (D.K.); jstriskovic@fizika.unios.hr (J.S.)}

\abstract{High energy photons from astrophysical sources are unique probes for some predictions of candidate theories of \ac{QG}. In particular, \acp{IACT} are instruments optimised for astronomical observations in the energy range spanning from a few tens of GeV to $\sim$100\,TeV, which makes them excellent instruments to search for effects of \ac{QG}. In this article, we will review \ac{QG} effects which can be tested with \acp{IACT}, most notably the \ac{LIV} and its consequences. It is often represented and modelled with photon dispersion relation modified by introducing energy-dependent terms. We will describe the analysis methods employed in the different studies, allowing for careful discussion and comparison of the results obtained with \acp{IACT} for more than two decades. Loosely following historical development of the field, we will observe how the analysis methods were refined and improved over time, and analyse why some studies were more sensitive than others. Finally, we will discuss the future of the field, presenting ideas for improving the analysis sensitivity and directions in which the research could develop.}


\begin{document}


\maketitle
\noindent\textbf{Content}

\contentsline {section}{\numberline {1}	Introduction and Motivation}{\pageref*{Sec:Introduction}}{section.1}%
\contentsline {subsection}{\numberline {1.1}A Proposal to Probe Quantum Gravity}{\pageref*{Subsec:Proposal}}{subsection.1.1}%
\contentsline {subsection}{\numberline {1.2}Modified Photon Dispersion Relation}{\pageref*{Subsec:ModifiedDispRel}}{subsection.1.2}%
\contentsline {section}{\numberline {2}Testing Energy-Dependent Photon Group Velocity}{\pageref*{Sec:TOF}}{section.2}%
\contentsline {subsection}{\numberline {2.1}The First Test with an Imaging Atmospheric Cherenkov telescope}{\pageref*{Subsec:Mrk421_Whipple}}{subsection.2.1}%
\contentsline {subsection}{\numberline {2.2}Fastest Variability in Blazars}{\pageref*{Subsec:Mrk501_MAGIC}}{subsection.2.2}%
\contentsline {subsection}{\numberline {2.3}Introducing the Maximum Likelihood Method}{\pageref*{Subsec:MaximumLikelihood}}{subsection.2.3}%
\contentsline {subsection}{\numberline {2.4}Results from the Maximum Likelihood Method on the Mrk\nobreakspace {}501 Flare from 2005}{\pageref*{Subsec:MLonMrk501_MAGIC}}{subsection.2.4}%
\contentsline {subsection}{\numberline {2.5}Sensitivity to the Lorentz Invariance Violation Effects}{\pageref*{Subsec:Sensitivity}}{subsection.2.5}%
\contentsline {subsection}{\numberline {2.6}Lorentz Invariance Violation Study on the Most Variable Blazar Flare}{\pageref*{Subsec:PKS2155_HESS}}{subsection.2.6}%
\contentsline {subsection}{\numberline {2.7}Extending to Higher Redshifts}{\pageref*{Subsec:PG1553_HESS}}{subsection.2.7}%
\contentsline {subsection}{\numberline {2.8}Exploring Lower Time Variability with the Crab Pulsar Observations by VERITAS}{\pageref*{Subsec:Crab_VERITAS}}{subsection.2.8}%
\contentsline {subsection}{\numberline {2.9}Applying the Maximum Likelihood Method to the Crab Pulsar with MAGIC}{\pageref*{Subsec:Crab_MAGIC}}{subsection.2.9}%
\contentsline {subsection}{\numberline {2.10}Lorentz Invariance Violation Study on a New Vela Pulsar}{\pageref*{Subsec:Vela_HESS}}{subsection.2.10}%
\contentsline {subsection}{\numberline {2.11}First Parallel Study of Energy-Dependent Photon Group Velocity and Gamma-ray Absorption on the Same Data Sample}{\pageref*{Subsec:Mrk501_HESS}}{subsection.2.11}%
\contentsline {subsection}{\numberline {2.12}First Lorentz Invariance Violation Study on a Gamma-Ray Burst Observed with Imaging Atmospheric Cherenkov Telescopes}{\pageref*{Subsec:GRB190114C}}{subsection.2.12}%
\contentsline {subsection}{\numberline {2.13}Lorentz Invariance Violation on \textit {Fermi}-LAT Gamma-Ray Bursts}{\pageref*{Subsec:Fermi-LAT}}{subsection.2.13}%
\contentsline {section}{\numberline {3}Modified Photon Interactions}{\pageref*{Sec:Modiefied_photon_interactions}}{section.3}%
\contentsline {subsection}{\numberline {3.1}Testing Lorentz Invariance Violation with Universe Transparency}{\pageref*{Subsec:UniverseTransparency}}{subsection.3.1}%
\contentsline {subsubsection}{\numberline {3.1.1}Influence of Lorentz Invariance Violation on Universe Transparency}{\pageref*{Subsubsec:when_it_all_began}}{subsubsection.3.1.1}%
\contentsline {subsubsection}{\numberline {3.1.2}Testing Lorentz Invariance Violation on Universe Transparency}{\pageref*{Subsubsec:Biteau_Williams}}{subsubsection.3.1.2}%
\contentsline {subsubsection}{\numberline {3.1.3}The Most Constraining Limits Based on Single Source Analysis}{\pageref*{Subsubsec:HESS_spectral}}{subsubsection.3.1.3}%
\contentsline {subsubsection}{\numberline {3.1.4}On How the Most Constraining Limits Were Obtained}{\pageref*{Subsubsec:Lang_et_al}}{subsubsection.3.1.4}%
\contentsline {subsection}{\numberline {3.2}Constraints on Violation of Lorentz Invariance from Atmospheric Showers Initiated by Multi-TeV Photons}{\pageref*{Subsec:LIVonShowers}}{subsection.3.2}%
\contentsline {subsection}{\numberline {3.3}Constraints on Lorentz Invariance Violation Based on Photon Stability}{\pageref*{Subsec:PhotonDecay}}{subsection.3.3}%
\contentsline {section}{\numberline {4}Summary and Discussion}{\pageref*{Sec:ResultsComparison}}{section.4}%
\contentsline {section}{\numberline {5}An Eye on the Future}{\pageref*{Sec:Future}}{section.5}%
\contentsline {subsection}{\numberline {5.1}Refinement of the Analysis Technique}{\pageref*{Subsec:AnalysisRefinement}}{subsection.5.1}%
\contentsline {subsection}{\numberline {5.2}Combining Data from Different Sources and Instruments}{\pageref*{Subsec:CombiningIACTs}}{subsection.5.2}%
\contentsline {subsection}{\numberline {5.3}Additional and Alternative Lorentz Invariance Violation Effects and Related Phenomena}{\pageref*{Subsec:OtherLIVeffects}}{subsection.5.3}%
\contentsline {section}{\numberline {6}Conclusions}{\pageref*{Sec:Conclusions}}{section.6}%
\contentsline {section}{Abbreviations}{\pageref{Sec:Abbreviations}}{}%
\contentsline {section}{References}{\pageref{Sec:References}}{}%

\section{Introduction and Motivation}
\label{Sec:Introduction}

The general theory of relativity is a beautiful and elegant theory, which connects the local matter and energy content to the curvature of spacetime, thus giving a classical description of gravity. It has been heavily tested and scrutinised ever since Albert Einstein proposed it in 1915 \cite{Einstein:1916vd}. Nevertheless, it breaks down in extreme circumstances such as singularities within black holes, or the early universe. Therefore, it is expected that there exists a more fundamental quantum theory of gravity, which can handle these extreme situations. Furthermore, a quantum theory of gravity would be a giant leap towards unification of all fundamental forces.

Theoretical endeavours in formulating the theory of \ac{QG} have explored different avenues (see, e.g., \cite{Rovelli:1989za, Rovelli:2007uwt, Oriti:2001qu, Niedermaier:2006wt, Ambjorn:2012jv, Horowitz:2004rn}). 
However, despite significant efforts, a complete and consistent description of gravity on a quantum level remains unknown. In addition, many of the \ac{QG} models include departures form the Lorentz symmetry (see, e.g., \cite{Kostelecky:1988zi,Gambini:1998it,Carroll:2001ws,Douglas:2001ba}). 
Performing measurements in the expected realm of \ac{QG} would strongly hint in which direction theoretical research should proceed. Unfortunately, the expected domain of \ac{QG} is the Planck scale
\footnote{Planck energy $E_{\mathrm{Pl}} = \sqrt{\hbar c^5/G} \approx 1.22 \times 10^{19}$\,GeV, Planck length $L_{\mathrm{Pl}} = \sqrt{\hbar G/c^3} \approx 1.62 \times 10^{-35}$\,m, Planck time $t_\mathrm{Pl} = \sqrt{\hbar G/c^5} \approx 5.39 \times 10^{-44}$\,s.}.
Even if \ac{QG} emerges at energies several orders of magnitude below $E_{\mathrm{Pl}}$, it is still vastly above the highest energies accessible in contemporary human-built accelerators.
When technology falls short, we turn to nature's own accelerators: active galactic nuclei, gamma-ray bursts, supernova remnants, pulsars, etc.
The most energetic particles detected up to date, a cosmic ray at $\sim$$3.2 \times 10^{11}$\,GeV \cite{Bird:1994uy, Halzen1995:CRlimit}, a neutrino at $\sim$$6.3 \times 10^{6}$\,GeV \cite{IceCube:2021rpz}, and a gamma ray at $\sim$$1.4 \times 10^{6}$\,GeV \cite{Cao2021}, while reaching energies higher than those achievable in Earth-based accelerators, are still more than a little shy of $E_{\mathrm{Pl}}$.
So what allows us to hope that we will measure an effect of \ac{QG}?

\subsection{A Proposal to Probe Quantum Gravity}\label{Subsec:Proposal}
In 1997, the distance of a \ac{GRB}\footnote{For a description of \acp{GRB} and an overview of observations, we refer the reader to Nava, L. Evolution of GRB observations over the past 30 years, to be printed in this Special Issue,} was measured for the first time. 
Indeed, following the detection of GRB~970508 \cite{1997IAUC.6649....1C}, an optical counterpart was observed~\cite{1997Natur.387..876D}, which allowed estimating of its redshift to $0.835 \le z \lesssim 2.3$ \cite{1997Natur.387..878M}. 
A strong flux of gamma rays from a quickly varying source detected at a cosmological distance incited {Amelino-Camelia~et~al.}~\cite{AmelinoCamelia:1997gz} to suggest that the signal from \acp{GRB} could be used to probe the structure of spacetime. 
The proposal was based on the idea of spacetime as a dynamical medium, which experiences quantum fluctuations due to \ac{QG} effects. While the scale of fluctuations was expected to be comparable to the Planck units, propagation of photons of substantially smaller energies could still be affected by it. Probing the fluctuations would result in an energy-dependent propagation speed, similar to what visible light experiences when propagating through a medium such as water or air.

\subsection{Modified Photon Dispersion Relation}\label{Subsec:ModifiedDispRel}
This behaviour can be modelled by modifying the standard photon dispersion relation in the following way:

\begin{equation}\label{eq:moddispastro}
    E^{2} = p^{2} c^{2} \times\left[1+\sum_{n=1}^{\infty}S_n\left(\frac{E}{E_{\mathrm{QG,}n}}\right)^{n}\right],
\end{equation}
where $E$ and $p$ are respectively the energy and momentum of a photon, $c$ is the Lorentz invariant speed of light, and the $E_{\mathrm{QG,}n}$ are the energy scales at which effects of \ac{QG} become significant. 
We will start discussing the values of $E_{\mathrm{QG,}n}$ in a short while, for now let us just acknowledge that $E/E_{\mathrm{QG,}n} \ll 1$ even for the most energetic gamma rays. Different modifying terms in the dispersion relation contribute less and less with increasing $n$. Therefore, usually, only the first two leading terms ($n=1$ or $n=2$) of the series are considered and independently tested for\footnote{Non-integer \cite{Calcagni:2016zqv}, as well as $n=-1$ and $n=0$ \cite{Colladay:1998fq, Carroll:1989vb, Kostelecky:2001mb} can also be considered. However, in gamma rays, the effect will be most pronounced for small, positive, integer values, so the focus is usually on $n=1$ and $n=2$.}. They are often referred to as linear and quadratic energy-dependent contributions, respectively. 
Letting $E_{\mathrm{QG,}n} \rightarrow \infty$ for all $n$ leads to the well-known Lorentz invariant photon dispersion relation. 
Parameters $S_n$ can take values $\pm1$, and their role will become apparent immediately. 
From Equation~(\ref{eq:moddispastro}) the energy-dependent photon group velocity can be easily derived as:

\begin{equation}\label{eq:photonvelocity}
    v_\gamma = \frac{\partial E}{\partial p} \simeq c \left[1+\sum_{n=1}^{\infty}S_n\frac{n+1}{2}\left(\frac{E}{E_{\mathrm{QG,}n}}\right)^{n}\right].
\end{equation}

Considering each modifying term independently, one can see that for $S_n=+1$, the velocity becomes greater than $c$, while for $S_n=-1$ it becomes smaller than $c$. These two behaviours are known as superluminal and subluminal, respectively.

Once a modification is introduced in the dispersion relation, various effects (other than changes in the photon speed) are conceivable, such as the modification of the electromagnetic interaction.
But whatever the effects of modifying the dispersion relation may be, they are minuscule because of the ratio $E/E_{\mathrm{QG,}n}$. 
The good news, however, is that the effects are cumulative. This is extremely important because gamma rays from some astrophysical sources take billions of years to reach Earth, allowing for these potential effects to accumulate, thus giving hope that we might be able to measure their consequences from~Earth.

Additionally, the effects of modifying the dispersion relation are more pronounced for higher energy photons. Thus, searching for them with \acp{IACT}, which are instruments optimised for astronomical observations in the \ac{VHE} gamma-ray band (100\,GeV\,$<E<$\,100\,TeV), is a sensible thing to do. Given their large collection area and good sensitivity, \acp{IACT} are excellent instruments for testing effects of \ac{QG} on gamma rays, and will be the main focus of this review. At lower energies, satellite-born detectors such as the \textit{Fermi}-\ac{LAT} benefit from more distant observations, but suffer from their lower effective area (see Sections~\ref{Subsec:Fermi-LAT} and \ref{Sec:ResultsComparison} for a brief comparison with the \ac{IACT} results). On the other hand, at higher energies water Cherenkov detectors such as \ac{HAWC} or \ac{LHAASO} have an advantage of observing in a higher energy range than \acp{IACT}. However, due to the rapid decrease of the flux of gamma rays at these energies they are handicapped by smaller statistics, which makes them less sensitive to fast flux variations (see Sections~\ref{Subsec:PhotonDecay} and \ref{Sec:ResultsComparison} for a brief comparison with the \ac{IACT} results).

Before taking a dive into the methods and results of probing \ac{QG} with \acp{IACT}, let us acknowledge that the modified photon dispersion relation is the usual starting point of experimental tests of \ac{QG} on gamma rays.
While some \ac{QG} models indeed do not preserve Lorentz symmetry, it is important to note that Equation~(\ref{eq:moddispastro}) is not a direct consequence of any particular \ac{QG} model. Given that there is no fully formulated theory of \ac{QG}, it would be overambitious to expect exact predictions. Rather, the modified dispersion relation can be regarded as a simple way of parameterizing and modelling phenomena not predicted by the current physical theories and laws. It, therefore, enables us to experimentally search for effects of those phenomena.

That being said, there are two main ways of modifying the dispersion relation that are usually considered. \ac{LIV}, the main focus of this review, implies the existence of a preferred inertial frame of reference, which breaks the Lorentz symmetry \cite{Colladay:1998fq}. However, there are also ways of modifying the photon dispersion relation, while at the same time preserving the Lorentz symmetry. One example is the so-called \ac{DSR} \cite{AmelinoCamelia:2002wr,AmelinoCamelia:2010pd}. In this model, the symmetry is deformed, rather than broken, and there is no preferred inertial frame of reference. 
Moreover, in order to keep the conservation laws covariant with respect to deformed symmetries of \ac{DSR}, the conservation laws themselves need to be modified. 
This fundamental difference between \ac{LIV} and \ac{DSR} becomes important when different possible effects of \ac{QG} are discussed. 
In particular, the kinematics, and possibly the dynamics, of electromagnetic interactions in the \ac{LIV} framework will differ from the Lorentz invariant ones. In the \ac{DSR} framework, on the other hand, the descriptions of interactions will be the same as (or only slightly different from) the Lorentz invariant descriptions. 
\ac{DSR} is a recently discussed promising avenue of research gaining attention and traction. However, there has been no published results from \acp{IACT} mentioning explicitly \ac{DSR} effects thus far. Therefore, in the rest of the text, we will, refer to all effects as \ac{LIV}, regardless of their true origin. We will, however, keep using $E_{\mathrm{QG,}n}$ to note the energy scales at which the effects become relevant. 
The details of either of these models, and their differences are out of the scope of this work. An interested reader is referred to a review paper by the COST~Action~18108\footnote{COST Action CA18108: Quantum gravity phenomenology in the multi-messenger approach (QG-MM, \href{https://qg-mm.unizar.es/}{https://qg-mm.unizar.es/}, accessed 15 July 2021) gathers researchers working on the theoretical and phenomenological predictions, and experimental searches for physical phenomena characteristic for of~\ac{QG}.} (in preparation) and references therein.

In this paper, we will focus on searches for signatures of \ac{LIV} in measurements with \acp{IACT}. We will discuss various effects of modifying the photon dispersion relation and their respective probes, adopting a chronological course. However, it is our intention (instead of simply recalling the most important studies performed) to analyse the evolution of the field, with a particular focus on the development of the analysis methods. Hopefully, this approach will inspire the authors and the readers alike to formulate new ideas on how to search for the effects of \ac{QG}, and pave the path for future research.
Historically, the first effect to be tested was the energy dependence of the photon group velocity, so results of different measurements of the photon time of flight will be covered first, in Section~\ref{Sec:TOF}. 
As stated above, \ac{LIV} can affect the kinematics and dynamics of electromagnetic interactions. This other important class of effects will be discussed in Section~\ref{Sec:Modiefied_photon_interactions}. 
We might as well break the suspense and state right away that no effects of \ac{QG} have been detected so far. Nevertheless, strong constraints have been set on the minimum value of the \ac{LIV} energy scale. These are usually expressed as lower limits at the 95\% confidence level. The results of different effects, obtained from various experiments and analysis methods will be mutually compared and their differences discussed in Section~\ref{Sec:ResultsComparison}. Finally, we turn towards the future in Section~\ref{Sec:Future}, to discuss opportunities for development and progress of this field of research.

\section{Testing Energy-Dependent Photon Group Velocity}
\label{Sec:TOF}

Assuming energy-dependent propagation speeds, two photons of energies $E_{\mathrm{2}} > E_{\mathrm{1}}$ emitted from a source at the same time will have different, energy-dependent, times of flight $t'_2$ and $t'_1$ respectively, finally reaching Earth with an energy-dependent time~delay~\cite{Jacob:2008bw}: 

\begin{equation}\label{eq:TimeDelay1}
    \Delta t' = t'_2 - t'_1 = t \frac{\Delta v_{\gamma}}{c} \simeq - S_n  \frac{n+1}{2}  \frac{E_\mathrm{2}^{n} - E_\mathrm{1}^{n}}{E_{\mathrm{QG,}n}^{n}}  D_{n}(z_{\mathrm{s}}),
\end{equation}
where $t$ is the time needed for a photon travelling with speed $c$ to reach the Earth\footnote{Note that the time delay can be both positive or negative, depending on whether the behaviour is subluminal or superluminal, respectively. According to the usual convention, a time delay is positive for subluminal behaviour, i.e., photon of a higher energy propagating at a lower speed than a lower energy photon.}. The time delay is proportional to a source distance parameter:

\begin{equation}\label{eq:LIVComovingDistance}
    D_{n}(z_{\mathrm{s}}) = \frac{1}{H_0} \bigintss_0^{z_{\mathrm{s}}} \frac{(1+z)^{n}}{\sqrt{\Omega_{\mathrm{m}}\left(1+z\right)^{3}+\Omega_{\Lambda}}} dz,
\end{equation}
where $z_{\mathrm{s}}$ is the source redshift, and $H_0$, $\Omega_{\mathrm{m}}$, and $\Omega_{\Lambda}$ represent cosmological parameters, respectively: the Hubble constant, the matter density parameter, and the dark-energy density parameter\footnote{Various \ac{LIV} studies use different values for cosmological parameters. However, given the precision of these studies, their final results are not strongly affected by the differences in the values of the cosmological parameters used, so we will treat them equally in that respect.}. 
The time delay expression was derived from comoving trajectories of particles, starting from their modified dispersion relations. More general and alternative expressions can be obtained by modifying the general relativistic dispersion relation as was done in \cite{Pfeifer:2018pty}, or by adopting that the spacetime translations are modified alongside with the modification of the dispersion relation \cite{Rosati:2015pga}.

\subsection{The First Test with an Imaging Atmospheric Cherenkov Telescope}
\label{Subsec:Mrk421_Whipple}
Soon after it was proposed that \acp{GRB} could be used to search for effects of \ac{LIV}, the first test using data from \acp{IACT} was performed. In fact, researchers observing with the Whipple telescope\footnote{The Whipple telescope (\href{https://veritas.sao.arizona.edu/whipple}{https://veritas.sao.arizona.edu/whipple}, accessed 15 July 2021) was the first \ac{IACT}. It consisted of a single 10\,m reflector dish. In operation from 1968 until 2013, it detected the very first TeV gamma-ray source, the Crab nebula \cite{Weekes:1989tc}, and the first \acs{AGN} detected in the same energy range Mrk~421~\cite{1992Natur.358..477P}.} already had a suitable data set available. Albeit, the source was not a \ac{GRB}, but the very first \ac{AGN} ever detected in the \ac{VHE} gamma-ray band, Markarian~421 (Mrk~421, redshift $z=0.031$) \cite{1992Natur.358..477P}. On 15 May 1996, Whipple observed the most rapid flare from Mrk~421 up to that time, with the flux doubling time of less than 15 minutes and including photons of energies up to several TeV \cite{1996Natur.383..319G}. This groundbreaking study used a rather rudimentary analysis: the data set was split in two energy bands ($E<1$\,TeV and $E>2$\,TeV) \cite{1999PhRvL..83.2108B}. In each energy band, the events were further subdivided in time bins of 280\,s. The distribution of arrival times of photons with energies $E>2$\,TeV was compared to the distribution of arrival times of events below 1\,TeV. The authors used the likelihood-ratio test\footnote{Not to be confused with the \acl{ML} (\acs{ML}) method presented in Section~\ref{Subsec:MaximumLikelihood}.} to compare the contents of time bins in the two energy ranges. 
In this study, no distinction was made between the subluminal and the superluminal behaviour. No delay in either direction was detected at the 95\% confidence level. Combined with the distance to Mrk~421, this result was translated into a lower limit on the \ac{LIV} energy scale $E_{\mathrm{QG,1}} > 4\times10^{16}$\,GeV. Only the linear contribution was considered in this first study.

\subsection{\label{Subsec:Mrk501_MAGIC}Fastest Variability in Blazars}

The observation of Mrk~421 with the Whipple telescope drew attention to flaring blazars as possible probes of \ac{LIV}. 
In the summer of 2005, the \ac{MAGIC} telescopes\footnote{\ac{MAGIC} (\cite{2016APh....72...61A, 2016APh....72...76A}, \href{https://magic.mpp.mpg.de/}{https://magic.mpp.mpg.de/}, accessed 15 July 2021) is a system of two semi-identical 17\,m reflector dish \ac{IACT}s. Located in the Roque de los Muchachos Observatory in the Canary island of La Palma, it has been in operation since 2004, first as a single MAGIC-I telescope. MAGIC-II was commissioned in 2009. Since then, \ac{MAGIC} has been observing as a stereoscopic telescope system.} observed two flares from the \ac{AGN} Markarian~501 (Mrk~501, redshift $z=0.034$) \cite{Albert:2007zd}. 
The data analysis revealed that the flux doubled in only 2 minutes, which remains until today the fastest flux variability ever observed from a blazar in the \ac{VHE} gamma-ray band. 
With the highest energies reaching $\sim$$10$\,TeV, the flux varying by an order of magnitude, 
it was a chance not to be missed. Moreover, there was an indication of a $4 \pm 1$\,min delay between the peaks in the light curves in the lowest (0.15--0.25\,TeV) and the highest (1.2--10\,TeV) energy bins on the 9th of July. 
A search for an energy-dependent photon time of flight in the Mrk~501 flare of this night was performed employing two distinct statistical analysis methods which we will now describe.

\textbf{\Ac{ECF}}\label{Subsubsec:ECF} method utilises the fact that a signal pulse propagating through a dispersive medium will be diluted, and its power (total energy per unit time), consequently, decreased (see, e.g., Section 7.9 in \cite{jackson1999classical}).
In the case of the Mrk~501 flare~\cite{2008PhLB..668..253M}, the data sample was chosen by selecting the most active part of the flare, i.e., the time interval in which the temporal distribution of events differs the most from a uniform distribution. We will mark the beginning and the end of this time interval as $t_1$ and $t_2$, respectively. 
The power of the signal was calculated as the sum of the energies of all the photons within the interval divided by the duration of the interval. Had the photons experienced any energy-dependent time delay, the power would have been smaller than without dispersion. One can then search for the maximal possible power by applying dispersion in the opposite direction, assuming different values of the \ac{LIV} energy scale. Specifically, in order to compute a new signal power, a new arrival time $t'_{\mathrm{i}}$ was calculated for each photon in the sample for a particular value of $E_{\mathrm{QG,}n}$:

\begin{equation}\label{eq:t_prime}
    t'_{\mathrm{i}} = t_{\mathrm{i}} + \eta_{n}  E_{\mathrm{i}}^{{n}},
\end{equation}
where $t_{\mathrm{i}}$ and $E_{\mathrm{i}}$ are, respectively, the measured arrival time and the measured energy of the $i$-th photon. Given the large values of $E_{\mathrm{QG,}n}$, various parameters are often introduced to facilitate numerical computations. Here the parameters $\eta_{{n}}$ is defined from Equation~(\ref{eq:TimeDelay1}) as:

\begin{equation}\label{eq:eta_n}
    \eta_n = - S_n  \frac{n+1}{2}  \frac{1}{E_{\mathrm{QG,}n}^{n}}  D_{n}(z_{\mathrm{s}}).
\end{equation}

This parameter is introduced for computational reasons because $\eta_n$ is $\mathcal{O}(1)$. Usually expressed in units of [s/GeV] (for $n=1$) or [s/GeV$^2$] (for $n=2$), $\eta_{{n}}$ indicates how much a photon will be delayed in arrival compared to a photon propagating at $c$ per every GeV of its energy. Limits on $E_{\mathrm{QG,}n}$ are then derived by inverting Equation~(\ref{eq:eta_n}).

The arrival time recalculation as described in Equation~(\ref{eq:t_prime}) was performed for each individual gamma ray, and only photons whose recalculated arrival times fell in the $[t_1, t_2]$ time interval were retained. In this way, an alternative sample of photons was constituted, and its total energy calculated. 
The procedure was repeated for different values of $\eta_{n}$ (i.e., different values of $E_{\mathrm{QG,}n}$). 
The \ac{ECF} was defined as the total energy as a function of $\eta_{n}$. The value of $\eta_{n}$ which maximises the \ac{ECF}, would recover the maximal signal power. In other words, it would correspond to the measurement of the dispersion which the gamma rays experienced because of the \ac{LIV} effects, assuming no other effects play a significant role.
The sensitivity of the method and the confidence interval for parameters $\eta_{n}$ were estimated using Monte Carlo simulations of the observed signal. Next, 1000 of simulated data sets were generated, and the \ac{ECF} method was applied to each of them. The most probable value of $\eta_{n}$ and its confidence interval were estimated from the distribution of $\eta_{n}$, which were then translated into lower limit on the \ac{LIV} energy scale at the 95\% confidence level. This particular analysis yielded $E_{\mathrm{QG,1}} > 2.1 \times 10^{17}$\,GeV, which was more constraining than the Whipple result on Mrk~421 flare from 1996 \cite{1999PhRvL..83.2108B} by an order of magnitude. In addition, for the first time the quadratic contribution was constrained, setting $E_{\mathrm{QG,2}} > 2.6 \times 10^{10}$\,GeV. 
Unlike the approach used for the Mrk~421 data analysis, the \ac{ECF} allowed for testing of superluminal, as well as subluminal behaviours. Nevertheless, only the subluminal behaviour was investigated.

There are several methods based on the idea of removing the dispersion from the data: the \ac{SMM} \cite{Vasileiou:2013vra, deAlmeida:2012cx}, the \ac{DisCan}~\cite{2008ApJ...673..972S}, and the \ac{MD}~\cite{Ellis:2008fc}; the main difference between these approaches being the way the sharpness of the light curve is quantified. We will investigate in more details a variation of the \ac{DisCan} method in Section~\ref{Subsec:Crab_VERITAS} and the \ac{SMM} in Section~\ref{Subsec:Fermi-LAT}.

\subsection{\label{Subsec:MaximumLikelihood}Introducing the Maximum Likelihood Method}

Originally, the \textbf{\ac{ML}} method was proposed for the analysis of the Mrk~501 data set described in the previous section. However, as we shall soon see, it became the standard analysis method used for searches of energy-dependent gamma-ray group velocity in \acp{IACT} data. Therefore, we will dedicate a separate section to its description. 
Introduced by {Mart\'{i}nez \& Errando} in \cite{2009APh....31..226M}, the authors argued that the analysis methods used should be unbinned (unlike the one previously employed in the case of Mrk~421 \cite{1999PhRvL..83.2108B}), in order to fully exploit the information carried by a relatively small gamma-ray sample. The \ac{ECF} method, used in \cite{2008PhLB..668..253M}, was indeed unbinned, however, it depended upon identifying and isolating the flares from the rest of the light curve. While this particular Mrk~501 light curve from 2005 had a relatively simple structure \cite{Albert:2007zd}, it was already recognised that the \ac{ECF} method would not be suited for the analysis of complex light curves, or segments of flares.
Therefore, the unbinned \ac{ML} method soon became a standard approach in searches for energy-dependent time delays, with every new study incorporating additional features and improvements. Here we will depart from the historical course, and describe the \ac{ML} method in its present form. 

In order to search for \ac{LIV}, the \ac{ML} method makes use of a profile likelihood ratio test:

\begin{equation}
    \label{Eq:likelihood_profile}
\lambda_p \left( \eta_n \mid \bm{\mathcal{D}} \right) = \frac{\mathcal{L} \left( \eta_n ; \bm{\hat{\hat{\nu}}} \mid \bm{\mathcal{D}} \right)}{\mathcal{L} \left( \widehat{\eta}_n ; \bm{\hat{\nu}} \mid \bm{\mathcal{D}} \right)} ,
\end{equation}
where $\eta_n$ is the \ac{LIV} parameter of order $n$ of interest, $\bm{\nu}$ represents the nuisance parameters, $\widehat{\eta}_n$ and $\bm{\hat{\nu}}$ are values that maximize the likelihood $\mathcal{L}$, $\bm{\hat{\hat{\nu}}}$ maximizes the likelihood $\mathcal{L}$ for a given $\eta_n$, and $\bm{\mathcal{D}}$ represents the observed data on which the analysis is performed. 
According to Wilks' theorem \cite{Wilks}, the distribution of $-2\ln \lambda_p(\eta_n|\bm{\mathcal{D}})$ follows a $\chi^2$ distribution with 1~degree of freedom for the true value of $\eta_n$, i.e., the one we are looking for. 
The 95\% confidence level one-sided upper limits are therefore derived by solving the following equation:

\begin{equation} \label{eqUL-1side}
    -2\ln \lambda_p(\eta^{\mathrm{UL}}|\bm{\mathcal{D}}) = 2.71,
\end{equation}
while 95\% confidence level two-sided upper limits are obtained using:

\begin{equation} \label{eqUL-2side}
    -2\ln \lambda_p(\eta^{\mathrm{UL}}|\bm{\mathcal{D}}) = 3.84.
\end{equation}

In the case where the conditions for Wilks' theorem are not fulfilled, one can calibrate intervals using Monte Carlo simulated samples of the null hypothesis.
The right value for any particular case can then be derived from the quantiles of the distribution of these simulations. For instance, the 95\% two-sided confidence interval is delimited by the lower and upper 2.5\% quantiles.

The likelihood function $\mathcal{L}$, for an observed number of events $N_{\mathrm{ON}}$, can be written as:
\begin{equation}\label{eq:Likelihood}
    \mathcal{L}(\eta_{{n}})= \prod_{i=1}^{N_{\mathrm{ON}}} \left(p^{(\mathrm{s})}_{i}  \frac{f^{(\mathrm{s})}(E_{i}, t_{i})}{\int_{E_{\mathrm{min}}}^{E_{\mathrm{max}}} dE \int_{t_{\mathrm{min}}}^{t_{\mathrm{max}}} f^{(\mathrm{s})}(E, t) dt} + p^{(\mathrm{b})}_{i} \frac{f^{(\mathrm{b})}(E_{i}, t_{i})}{\int_{E_{\mathrm{min}}}^{E_{\mathrm{max}}} dE \int_{t_{\mathrm{min}}}^{t_{\mathrm{max}}} f^{(\mathrm{b})}(E, t) dt}\right),
\end{equation}
where $f^{(\mathrm{s})}(E, t)$ represents the \ac{PDF} for observing a gamma ray of reconstructed energy $E$ at the moment $t$, while $f^{(\mathrm{b})}(E, t)$ is the \ac{PDF} for observing a background event of reconstructed energy $E$ at the moment $t$. The energies $E_i$ are bounded by $E_{\mathrm{min}}$ and $E_{\mathrm{max}}$, respectively the minimum and maximum energy considered in the analysis expressed in reconstructed (i.e., measured) energy, which in turn usually depend on the instrument and observation conditions. Similarly the times $t_i$ are bounded by $t_{\mathrm{min}}$ and $t_{\mathrm{max}}$. These four quantities are used to compute the normalisation factors of both the signal and the background part of the likelihood function. Additionally, in standard \ac{IACT} analyses, the so-called ON region in the field of view contains both signal and background events. Therefore, $p^{(\mathrm{s})}_{i}$ and $p^{(\mathrm{b})}_{i}$ are the probabilities for the event $i$ to belong to the signal or the background, respectively.
The \ac{PDF} for observing a gamma ray of reconstructed energy $E$ at the moment $t$

\begin{equation}\label{eq:PDF}
    f^{(\mathrm{s})}(E, t) = \int_0^{\infty} F(t + \eta_{n}E^{n}) \,  \Phi_{\mathrm{obs}}(E) \, G \left( E, E_{\mathrm{true}} \right) \, A_{\mathrm{eff}}(E_{\mathrm{true}}, t) \, dE_{\mathrm{true}},
\end{equation}
contains all available information about the emitted signal at the source, the gamma-ray propagation effects, and the detection process. Namely,
\begin{itemize}\label{item:ML-PDF}
    \item function $F(t)$ is the observed light curve. Here, by taking $F(t + \eta_{n}E^n)$, it is ``corrected'' for the potential time delay induced by the \ac{LIV} effects. In this way, assuming that individual events suffered an energy-dependent time delay, and that no other dispersion effects were present, one obtains a source-intrinsic light curve, often referred to as a light curve template. In practice, there are different ways of obtaining $F(t)$.  
    \item $\Phi_{\mathrm{obs}}(E)$ represents the observed spectral distribution of gamma rays. As it will be described in more details in Section~\ref{Subsec:UniverseTransparency}, $\Phi_{\mathrm{obs}}$ can be decomposed into a source intrinsic spectrum term and an absorption term. The latter usually implies the absorption of gamma rays on the \ac{EBL}, as discussed in Section~\ref{Subsec:UniverseTransparency}, but can easily accommodate any additional effect (or modification of this particular one) that can affect the spectral distribution of gamma rays during their propagation towards the detector. 
    \item $G \left( E, E_{\mathrm{true}} \right)$, contains the information about the energy resolution and the bias of the instrument. $E_{\mathrm{true}}$ is the true energy of a particular event, and $G \left( E, E_{\mathrm{true}} \right)$ is the \ac{PDF} of $E_{\mathrm{true}}$ being measured as $E$. 
    \item \textls[-12]{The final ingredient, $A_{\mathrm{eff}}(E_{\mathrm{true}}, t)$ represents the collection area (i.e., acceptance) of the instrument expressed in true energy $E_{\mathrm{true}}$. In the most general case, it can change with time, especially if the data were collected in different observation~conditions.} 
\end{itemize}

The \ac{PDF} for observing a background event of reconstructed energy $E_i$ at the moment $t_i$ has fundamentally the same form as the \ac{PDF} for signal events, see Equation~(\ref{eq:PDF}). However, the origin of background events is generally not known, so time of flights of individual events (whether affected by \ac{LIV} or not) cannot be determined. Therefore, both temporal and energy distributions of events are taken as observed on Earth. Concretely, in \mbox{Equation~(\ref{eq:PDF})}, when used for background events 
$\eta_n = 0$, and $F(t)$ and $\Phi_{\mathrm{obs}} (E)$ are the measured background light curve and spectrum on Earth. 
The final pieces of puzzle are probabilities for each event to be part of the signal or background, $p^{(\mathrm{s})}_{i}$ and $p^{(\mathrm{b})}_{i}$. In \acp{IACT}, the signal is estimated from a region around the source position in the field of view, usually referred to as the ON region. However, besides the signal, the ON region contains an irreducible contribution from the background. The background is estimated from the so-called OFF region, a region in the field of view which contains no sources of gamma rays, and is observed under the same conditions as the ON region. Usually, the probabilities for each event to be a part of the signal or background are calculated as follows:

\begin{equation}\label{eq:gammaProb}
    p^{(\mathrm{s})}_{i} = \frac{N_{\mathrm{ON}} - \alpha N_{\mathrm{OFF}}}{N_{\mathrm{ON}}},\qquad p^{(\mathrm{b})}_{i} = \frac{\alpha N_{\mathrm{OFF}}}{N_{\mathrm{ON}}}
\end{equation}
where $N_{\mathrm{ON}}$ is the total number of events in the ON region, $N_{\mathrm{OFF}}$ the total number of events in the OFF region, and $\alpha$ is the ratio of effective exposure times in the two: $\alpha=t_{\mathrm{ON}}/t_{\mathrm{OFF}}$\footnote{For a more detailed definition and discussion of $alpha$, we refer the interested reader to \cite{Berge:2006ae}.}. 
A legitimate objection to the \ac{ML} method is that it relies on our knowledge of source-intrinsic processes, which is limited at best. In that sense, the \ac{ECF} or similar methods like the \ac{SMM} (see Section~\ref{Subsec:Fermi-LAT}) have the advantage of not depending on our knowledge of source-intrinsic effects.
However, it is quite imaginable that there are source-intrinsic dispersive processes, which could mimic effects of \ac{LIV}\footnote{For more details on the modelling of these effects in the context of \ac{LIV}, an interested reader can refer to~\cite{Perennes:2019sjx}.}. These would not depend on the source redshift, and could be ``filtered out'' by considering sources at different redshifts. Nevertheless, combining the results of different analyses might prove tricky for \ac{ECF} and related methods. The likelihood function, on the other hand, should tackle that task with relative ease, as we will discuss in Section~\ref{Subsec:CombiningIACTs}. 

Another possible source of systematic effects are secondary gamma rays, which can be produced through one of the following processes: (i) hadrons accelerated within a source interact with the surrounding electromagnetic fields to produce neutral pions, which decay into gamma rays, (ii) gamma rays emitted from the source interact with magnetic fields to produce electron-positron pairs, which can create secondary gamma rays either through annihilation, or by inverse-Compton scattering of lower-energy photons. Secondary gamma rays could create a false signal, especially in the analysis methods based on individual events, such as time of flight studies. However, since secondary gamma rays are not produced within the observed source, their origin is not necessarily on the line of sight. A significant rate of secondary gamma rays would manifest as an extended emission, so-called halo, around an otherwise point-like source. Indeed, several studies searching for gamma-ray halos have been performed, but have shown no evidence thereof (see, e.g., \cite{MAGIC:2010goh,HESS:2014kkl,Fermi-LAT:2018jdy}, see also \cite{Batista:2021rgm} and references therein). Though an occasional secondary gamma ray might be mistakenly treated as the signal, the effect should be minor, and it would diminish with an increasing size of a data sample.

\subsection{\label{Subsec:MLonMrk501_MAGIC}Results from the Maximum Likelihood Method on the Mrk~501 Flare from 2005}

In this first application of the \ac{ML} method, the light curve of the Mrk~501 flare from 2005 was modelled with a Gaussian superimposed on top of a constant baseline emission from the source. A background contribution was not considered because of its negligible contribution in such a flare. 
The results showed $\eta_\mathrm{1}$ (see Equation~(\ref{eq:eta_n})) departing from zero by slightly more than $2\sigma$, implying an energy-dependent time delay \cite{2008PhLB..668..253M}. $\eta_\mathrm{2}$, on the other hand, was consistent with zero. Again, only the subluminal behaviour was tested for. 
The corresponding \ac{LIV} energy scales were $E_{\mathrm{QG,1}} = 0.30^{+0.24}_{-0.10}\times 10^{18}$\,GeV and $E_{\mathrm{QG,2}} = 0.57^{+0.75}_{-0.19}\times 10^{11}$\,GeV for the linear and quadratic contributions, respectively \cite{2009APh....31..226M}. 
{Mart\'{i}nez \& Errando} refrained from interpreting the results and focused on the description of the method. The nonzero time delay was instead discussed in \cite{2008PhLB..668..253M}. A possibility of a bias in the \ac{ML} analysis was investigated on a set of simulated Monte Carlo samples. 
An independent researcher simulated data sets with injected energy-dependent time delays. These data sets were blindly analysed using the \ac{ML} method, which correctly reconstructed the injected delay values.
It was concluded that the effect was real, although the statistical significance was too low to claim a discovery. Finally, it was concluded that the results obtained with the \ac{ECF} and the \ac{ML} methods were mutually consistent. Furthermore, some investigations of emission models suggest that the energy-dependent time delay could be a consequence of source intrinsic spectral variability in time, occurring either because of the acceleration of particles or the absorption of gamma rays \cite{Sitarek:2009za, Sitarek:2010ky}. 
In summary, the study on the Mrk~501 flare data from 2005, not only significantly tightened the constraints on the \ac{LIV} energy scale, compared to the pioneering study of Whipple \cite{1999PhRvL..83.2108B}, but it also motivated the introduction of a novel analysis method, and served as a cross check between two fundamentally different analysis approaches. This data set was also the first one to be studied with two fundamentally different analysis approaches, allowing comparisons between their results.

\subsection{Sensitivity to the Lorentz Invariance Violation Effects}
\label{Subsec:Sensitivity}

After taking a look at the first searches for the possible signatures of \ac{LIV} effects in \acp{IACT} data, this is a good place to analyse what properties a signal should have in order to be considered a good probe of such an effect.
As we have already discussed in the beginning of Section~\ref{Sec:TOF}, the more energetic photons will be more strongly affected by the \ac{LIV}. Therefore, sources with spectra extending to higher energies, and with comparatively larger population of higher energy photons (colloquially called ``harder spectra'', as opposed to ``softer spectra''), are more favourable. Furthermore, the farther the source, the more the effect will be accumulated. However, a large distance carries the caveat that \ac{VHE} gamma rays are partially absorbed on the \ac{EBL} (see the description of the \ac{ML} method in page~\pageref{item:ML-PDF} et sec., and Section~\ref{Subsec:UniverseTransparency}), which softens the spectra and depletes data samples of the most energetic photons. 
The time delay between two photons of different energies will also be more pronounced for smaller values of $E_{\mathrm{QG,}n}$, so smaller time delay means stronger constraint on the \ac{LIV} energy scale. However, it is entirely possible that there are emission time delays present within sources, which could mimic or conceal \ac{LIV}-induced arrival time delays.
Considering our limited knowledge of emission mechanisms, the emission times cannot be precisely modelled.
Instead, the emission time has to be constrained based on the flux variability timescale. Emission is more probable during periods of higher flux. However, high flux on its own is not enough. If the flux is constant, or changing monotonically, an application of a spectral dispersion, will not change the shape of the light curve. A variable light curve, on the other hand, will be smeared due to spectral dispersion. The effect will be more pronounced for stronger dispersion, and more detectable for faster changing flux.
By inverting Equation~(\ref{eq:TimeDelay1}), one can make a crude estimate of how \ac{LIV} energy scale depends on the highest energies of detected photons ($E_{\mathrm{max}}$), light curve variability timescale ($t_{\mathrm{var}}$), and the redshift of the source ($z_{\mathrm{s}}$). These dependencies are summarized in Table~\ref{Tab:TOFsensitivity}. Note that the power of the dependence on the redshift was numerically computed for $z_{\mathrm{s}}$ up to $\sim$$10$, which is much further than what current \acp{IACT} can probe.
\begin{specialtable}[H] 

    \tablesize{\small}
\caption{\label{Tab:TOFsensitivity}Dependence of $E_{\mathrm{QG,}n}$ on the characteristics of the source and the sample. $E_{\mathrm{max}}$ is the highest photon energy in the sample, $t_{\mathrm{var}}$ is the shortest variability timescale in the light curve, and $z_{\mathrm{s}}$ is the redshift of the source.}
\setlength{\cellWidtha}{\columnwidth/5-2\tabcolsep+0.0in}
\setlength{\cellWidthb}{\columnwidth/5-2\tabcolsep+0.0in}
\setlength{\cellWidthc}{\columnwidth/5-2\tabcolsep+0.0in}
\setlength{\cellWidthd}{\columnwidth/5-2\tabcolsep+0.0in}
\setlength{\cellWidthe}{\columnwidth/5-2\tabcolsep+0.0in}
\scalebox{1}[1]{\begin{tabularx}{\columnwidth}{
>{\PreserveBackslash\centering}m{\cellWidtha}
>{\PreserveBackslash\centering}m{\cellWidthb}
>{\PreserveBackslash\centering}m{\cellWidthc}
>{\PreserveBackslash\centering}m{\cellWidthd}
>{\PreserveBackslash\centering}m{\cellWidthe}}
\toprule
    $E_{\mathrm{QG,1}}$ & $\propto$ & $E_{\mathrm{max}}$ & $t_{\mathrm{var}}^{-1}$ & $z_{\mathrm{s}}^{\sim1}$\\
    \addlinespace[0.2cm]
    $E_{\mathrm{QG,2}}$ & $\propto$ & $E_{\mathrm{max}}$ & $t_{\mathrm{var}}^{-1/2}$ & $z_{\mathrm{s}}^{\sim2/3}$\\
\bottomrule
\end{tabularx}}
\end{specialtable}
 
There is another parameter, not present in Equation~(\ref{eq:TimeDelay1}), whose importance becomes apparent through the data analysis. That is the size of the sample. Its influence on $E_{\mathrm{QG,}n}$ depends on the analysis method, and is difficult to estimate it the way it was done for other parameters in Table~\ref{Tab:TOFsensitivity}. The general rule, though, is simple: the more the better. More specific estimates will be discussed on particular cases.

Based on this simple analysis, three types of sources are considered to be suitable for testing of \ac{LIV} on gamma rays: 

\textbf{Pulsars} can have rotation periods as short as a few milliseconds, although the ones detected with \acp{IACT} so far have periods of at least a few tens of milliseconds. The only four pulsars that have been detected with \acp{IACT} so far are the Crab pulsar~\cite{2008Sci...322.1221A}, the Vela pulsar~\cite{Abdalla:2018vik}, the Geminga pulsar~\cite{Acciari:2020zvo}, and PSR~B1706-44~\cite{Spir-Jacob:2020jth}. Their pulsation is highly regular, which makes it predictable, and allows stacking of signal from different periods, thus increasing the detected statistics. Additionally, these four pulsars are located in the Milky Way. This relatively close proximity significantly impairs the sensitivity of \ac{LIV} tests performed on pulsar data.

\textbf{Gamma-ray busts} are powerful transient cosmic explosions, usually associated with collapses of massive stars into black holes (long \acp{GRB}), or mergers of neutron stars (short \acp{GRB}). Their light curves are variable on timescales of a second. Unlike pulsars, \acp{GRB} are completely unpredictable. Satellite-borne detectors with a large field of view, such as \ac{GBM} \cite{Meegan:2009qu} and \ac{LAT} \cite{Atwood:2009ez} onboard satellite \textit{Fermi}\footnote{The Fermi Gamma-ray Space Telescope (\href{https://fermi.gsfc.nasa.gov/}{https://fermi.gsfc.nasa.gov/}, accessed 15 July 2021), formerly known as GLAST, was launched in 2008.} on average detect one \ac{GRB} almost every day \cite{Bhat:2016odd}. However, \acp{IACT} with a rather small field of view (an order of few degrees) rely on alerts from satellite borne detectors to trigger observations. Furthermore, because of their large distances, \ac{VHE} gamma rays are strongly absorbed on the \ac{EBL}. For these reasons, \acp{GRB} are elusive and notoriously difficult to detect with \acp{IACT}, with only four detected to date (\cite{Acciari:2019dxz, Arakawa:2019cfc, 2019GCN.25566....1D, 2020GCN.29075....1B}). However, due to their short variability timescales, combined with large distances, once a \ac{GRB} is detected, the signal becomes a valuable asset for probing \ac{QG} (see Section~\ref{Subsec:GRB190114C}).

\textbf{Active galactic nuclei} are persistent sources at distances comparable to \acp{GRB}. During their flaring states, they emit signals abundant in \ac{VHE} gamma rays, with flux variability timescales on the order of minutes. Although unpredictable, flares usually last longer than \acp{GRB}. In addition, they emit stronger fluxes, with the most energetic photons reaching higher energies. All of this makes flares from \ac{AGN} easier to detect with \acp{IACT} compared to \acp{GRB}.

\subsection{Lorentz Invariance Violation Study on the Most Variable Blazar Flare}
\label{Subsec:PKS2155_HESS}
While the Mrk~501 flare observed by \ac{MAGIC} showed the fastest changing gamma-ray flux in blazars, it had a rather simple structure. Almost exactly one year later, on 28 July 2006, while the \ac{LIV} data analysis on the Mrk~501 sample was still ongoing, another promising flare occurred. This time around it was the \ac{H.E.S.S.}\footnote{The \ac{H.E.S.S.} array (\cite{2020APh...11802425A}, \href{https://www.mpi-hd.mpg.de/hfm/HESS/}{https://www.mpi-hd.mpg.de/hfm/HESS/}, accessed 15 July 2021) is located in Khomas Highland plateau of Namibia. It consists of four 12\,m telescopes commissioned in 2004. In 2012, the array was extended with a 28\,m telescope, which marked the beginning of the \ac{H.E.S.S.}-II phase.} that observed a flare from blazar PKS~2155-304 \cite{Aharonian:2007ig}.
During an $\sim$$85$\,min observation, flare with a quite complex structure was detected, variable on the scale of $\sim$$200$\,s with several local minima and maxima, and with the signal to background ratio above 300. At the same time, no significant changes of spectrum were found.
The highest flux reached more than 15\,\ac{C.U.}\footnote{The Crab nebula is a pulsar wind nebula. It was the first source of gamma rays to be reliably detected with an \ac{IACT} \cite{Weekes:1989tc}. It is the brightest steady source of \ac{VHE} gamma rays, which which is why it is used as a standard candle in gamma astronomy. Gamma-ray flux is often expressed in units of Crab nebula flux (Crab units) in the same energy range.} above 200\,GeV, and a total of more than eleven thousand gamma rays were detected, reaching the highest energies of $\sim$$4$\,TeV. 
Moreover, PKS~2155-304 is located at a redshift of $z = 0.116$, more than three times larger than Mrk~421 and Mrk~501.

Several studies of energy-dependent time delay were performed using this signal.
The first one, published soon after the flare was observed, used two different statistical methods, both estimating time lag between light curves in different energy ranges \cite{Aharonian:2008kz}. 

\textbf{\Ac{MCCF}} was originally developed for timescale analysis of spectral lags, and it enables searches for time lags shorter than the temporal resolution of light curves \cite{Li:2004kp}. In this case, the data were split in two energy bins: 200--800\,GeV and $>$800\,GeV, and the \ac{MCCF} was used to estimate the time lag between the light curves in in these two energy ranges. The analysis resulted in the most stringent constraint on the linear contribution up to that time $E_{\mathrm{QG,1}} > 7.2\times10^{17}$\,GeV; more than two times stronger limit than the one set by \ac{MAGIC} on Mrk~501 data. The lower limit on the quadratic contribution, on the other hand, was set at $E_{\mathrm{QG,2}} > 1.4\times10^{9}$\,GeV; more than 40~times lower than the one from Mrk~501 by \ac{MAGIC} using the \ac{ML} method, and almost 20~times lower than the one set with the \ac{ECF}. 

\textbf{\Ac{CWT}} method relies on identifying extrema in two energy bands and measuring their relative time delay. In this case, the chosen energy ranges were: 210--250\,GeV and $>$600\,GeV, and two pairs of extrema were identified. Only the constraint on the linear term was set at $E_{\mathrm{QG,1}} > 5.2\times10^{17}$\,GeV, thus confirming the constraint obtained using the \ac{MCCF} method.

Relying on the rule of thumb, laid out in Table~\ref{Tab:TOFsensitivity}, it was expected that the larger distance and faster flux variability of PKS~2155-304, compared to Mrk~501, would make this study more sensitive to the linear modifying term of the dispersion relation. The influence of these two variables to the quadratic contribution is somewhat smaller, because of the exponents, allowing a stronger influence of the highest gamma-ray energies. Nevertheless, it seems unlikely that a factor of 2.5 difference in the highest energies alone would result in a factor of forty difference between limits on the quadratic contribution. It is more likely that the \ac{MCCF} and \ac{CWT} methods do not fully exploit all of the potentials of the PKS~2155-304 data sample. 

When proposing the \ac{ML} method in \cite{2009APh....31..226M}, the authors were already aware of the PKS~2155-304 flare, and decided to test their method on that signal as well. Since they did not have the access to the actual data set, and the method relied on individual events, they generated Monte Carlo simulated data sets, based on the published information on the PKS~2155-304 flare. It was estimated that the application of the \ac{ML} method on the PKS~2155-304 flare sample would be more than six times more sensitive to $E_{\mathrm{QG,}n}$ compared to the Mrk~501 case. Moreover, the authors analysed where the sixfold improvement came from, and came up with similar conclusions as we have just discussed: (i) the higher redshift contributed a factor of three, (ii) larger sample of PKS~2155-304, albeit with the highest energies lower than in the Mrk~501 sample, added another factor of two, and (iii) more complex light curve shape was responsible for an additional factor. However, the authors also noted that it was in fact the fastest single change of flux, i.e., the fastest rise time or fall time in the entire light curve, which dominated the sensitivity. 

Following the study by {Mart\'{i}nez \& Errando} \cite{2009APh....31..226M}, the \ac{H.E.S.S.} Collaboration performed another search for effects of \ac{LIV} in the PKS~2155-304 flare data, this time fully adopting the \ac{ML} method \cite{2011APh....34..738H}. 
For this occasion, a particular \ac{H.E.S.S.} data analysis was performed, focusing on the initial 4000\,s of the observation, during which both the flux and its variability were the highest. Upon applying some additional cuts on the data, only 3526 events remained (out of more than 11,000 in the original data set) in the 0.25--4.0\,TeV energy range. This resulted in a strong background suppression, and a very good fit of the light curve and the spectrum. Based on optimisation using Monte Carlo simulations, the data were finally separated in two energy bins: 0.25--0.28\,TeV and 0.3--4.0\,TeV. The lower bin was used to create the light curve template. The data were fitted with a sum of a constant baseline emission and five consecutive asymmetric Gaussian curves. The events from the higher energy bin were used to calculate the likelihood. 
The results were $E_{\mathrm{QG,1}} > 2.1 \times 10^{18}$\,GeV and $E_{\mathrm{QG,2}} > 6.4 \times 10^{10}$\,GeV for the linear and quadratic term, respectively, both significantly more constraining than the ones obtained in the previous analysis by \ac{H.E.S.S.} using \ac{MCCF}, demonstrating the dominance of the \ac{ML} method on a concrete case. Furthermore, both results were in line with the assessments by {Mart\'{i}nez \& Errando}, and finally, both were the most constraining lower limits on the \ac{LIV} energy scale up to that time. Discussing their results, the authors reached similar conclusions as {Mart\'{i}nez \& Errando} in their work. In particular, the higher sensitivity was due to the high flux variability and large data sample, while the lower maximal energies somewhat impaired the sensitivity. Furthermore, the uncertainty on the estimated parameter depended mostly on the width of the individual flux peaks, which was in agreement with the conclusion by {Mart\'{i}nez \& Errando} that the sensitivity is dominated by the fastest single change of flux. 
Final important point was that the estimated parameter uncertainty only mildly depended on the number of events used to calculate the likelihood, meaning that robust results are obtainable even with small data~sets.

\subsection{Extending to Higher Redshifts}
\label{Subsec:PG1553_HESS}

On 26 and 27 April 2012, the \ac{H.E.S.S.} telescopes observed a flare from the blazar PG~1553+113 \cite{Abramowski:2015ixa}. The flux was three times higher than the archival measurements, with an indication of intra-night variability. 
Interestingly, the redshift of the source had been only loosely constrained prior to this study. In order to estimate the redshift more precisely, the authors devised a method based on Bayesian statistics, which relies on accounting for the absorption of \ac{VHE} gamma rays on the \ac{EBL}\footnote{For details on the absorption of \ac{VHE} gamma rays on the \ac{EBL}, see Section~\ref{Subsec:UniverseTransparency}.}. This enabled them to estimate the redshift to be $z = 0.49 \pm 0.04$.
Though the flux showed only a hint of intra-night variability, the relatively large redshift encouraged the authors to perform a search for an energy-dependent time delay. Observations from the second day were used for that purpose. Unlike the flare from PKS~2155-304 (see Section~\ref{Subsec:PKS2155_HESS}) the signal to background ratio in this case was only 2. Due to this high background contamination, a \ac{PDF} for the background had to be introduced into the likelihood function for the first time. 
Events from the energy range 300--789\,GeV, the upper edge corresponding to the last significant bin, were used. The sample was separated into a lower energy bin used to create the light curve template, and a higher bin used for the \ac{ML} calculation. The delimiter between these two bins was set at 400\,GeV, approximately corresponding to the median of the sample.
The results, $E_{\mathrm{QG,1}} > 4.1\times10^{17}$\,GeV, $E_{\mathrm{QG,2}} > 2.1\times10^{10}$\,GeV for the subluminal scenario, and $E_{\mathrm{QG,1}} > 2.8\times10^{17}$\,GeV, $E_{\mathrm{QG,1}} > 1.7\times10^{10}$\,GeV for the superluminal scenario, did not further constrain the \ac{LIV} energy scale, but confirmed the already existing limits on the quadratic term. The bounds on the linear term were an order of magnitude below the ones set by \ac{H.E.S.S.} on PKS~2155-304 flare \cite{2011APh....34..738H}. 
The authors did not discuss the reasons for the lower sensitivity, however, referring again to our rule of thumb (Table~\ref{Tab:TOFsensitivity}), it seems safe to conclude that this study benefited from the high redshift of the source, while paying dues to the lower gamma ray energies detected, the modest sample size, and a marginal flux~variability.

\subsection{Exploring Lower Time Variability with the Crab Pulsar Observations by VERITAS}
\label{Subsec:Crab_VERITAS}

The idea of using pulsar emission to search for \ac{LIV} was first applied to Crab pulsar observations by EGRET~\cite{Kaaret:1999ve}. The Crab pulsar (PSR~J0534+2200) is located at the center of the Crab nebula at $2.0\pm0.5$\,kpc~\cite{Kaplan:2008qm} from Earth, and has a period of rotation of $\sim$$33$\,ms~\cite{Manchester:2004bp}. In 2011, the \ac{VERITAS} reported the observation of gamma-ray emission from the Crab pulsar above $100$\,GeV~\cite{2011Sci...334...69V}. Its phaseogram, i.e., its emission as a function of the pulsar rotational phase $\phi$, distinctly shows a main pulse (referred to as P1) and an inter-pulse (referred to as P2) at a phase $\phi$$\sim$$0.4$ from P1. For the \ac{LIV} analysis~\cite{2011ICRC....7..256O,2013ICRC...33.2768Z}, the authors made use of the \ac{PC} method. This method can be used to look for an average phase delay $\Delta \phi$ between photons from two different energy bands with mean energies $E_1$ and $E_2$ for the lower and the higher energy band, respectively:

\begin{equation}\label{eq:pulsar-peak-comparison}
    \Delta \phi_n = -S \frac{d_{\mathrm{Crab}}}{c  P_{\mathrm{Crab}}} \frac{n+1}{2} \frac{E_{2}^n - E_{1}^n}{E_{\mathrm{QG},n}^n}
\end{equation}
where $d_{\mathrm{Crab}}$ is the distance to the Crab pulsar, $P_{\mathrm{Crab}}$ its period and $c$ the Lorentz invariant in vaccuo speed of light. Note that the phase $\phi$ is a practical quantity when describing pulsar behavior, nevertheless since $\Delta t = \Delta \phi  P_{\mathrm{Crab}}$ one immediately recovers Equation~(\ref{eq:TimeDelay1}) from Equation~(\ref{eq:pulsar-peak-comparison}) under the assumption that $D_{n}(z_{\mathrm{s}}) \approx d$ which is true for such nearby sources as pulsars. The authors used this method to compare the mean fitted pulse position obtained with \ac{VERITAS} above $120$\,GeV to the one obtained with \textit{Fermi}-\ac{LAT} above $100$\,MeV~\cite{Abdo:2009ec}. The peak positions agreed within statistical uncertainties, therefore a 95\% confidence upper limit on their timing difference of $100\,\upmu\mathrm{s}$ could be derived. This limit was then converted into limits on $E_{\mathrm{QG_n}}$ by reversing Equation~(\ref{eq:pulsar-peak-comparison}):

\begin{equation}\label{eq:pulsar-peak-comparison-limit}
    E_{\mathrm{QG},n} > \left( -S \frac{d_{\mathrm{Crab}}}{c  P_{\mathrm{Crab}}} \frac{n+1}{2} \frac{E_{2}^n - E_{1}^n}{\Delta \phi_n} \right)^{\frac{1}{n}},
\end{equation}
yielding $E_{\mathrm{QG,1}} > 3.0\times 10^{17}$\,GeV and $E_{\mathrm{QG,2}} > 7.0\times 10^{9}$\,GeV in the subluminal scenario ($S=-1$). Note that in Equations~(\ref{eq:pulsar-peak-comparison}) and (\ref{eq:pulsar-peak-comparison-limit}), the distance parameter $D_{n}(z_{\mathrm{s}})$ from Equation~(\ref{eq:LIVComovingDistance}) was replaced by a more standard distance $d$ as pulsars are sources within the Milky Way, hence their distance is not properly described by the redshift. This means that the last column of Table~\ref{Tab:TOFsensitivity} is different in the case of pulsars, indeed $E_{\mathrm{QG},n}$ will be proportional to~$d^{1/n}$.

A variation of the \ac{DisCan} method was also used in this work~\cite{2013ICRC...33.2768Z}. It was first introduced in 2008~\cite{2008ApJ...673..972S} and, as its name suggests, consists in looking for the \ac{LIV} parameter that best cancels out any time dispersion in the data. As such this method is a variation of the \ac{ECF} with a different cost function. The variation consisted in the use of the $Z_{m}^{2}$ test~\cite{Buccheri:1983zz} as a test statistic (with $m=20$ resulting from Monte Carlo optimization for this particular case) applied to the phased data to look for the potential \ac{LIV} effect. This $Z_{20}^{2}$ \ac{DisCan} method yields a best value of $\eta_1 = -0.49 \, \upmu \mathrm{s/GeV}$. The calibration of the method using 1000 Monte Carlo simulations allowed the authors to establish that this value was only $1.4 \, \sigma$ away from the null hypothesis and therefore compatible with it. The 95\% confidence level limits on $\eta_1$ reached $-1.2 \, \upmu\mathrm{s/GeV}$ and $1.1 \, \upmu\mathrm{s/GeV}$ for the lower and upper limits, respectively. These results were then translated into the following limits: $E_{\mathrm{QG,1}} > 1.9\times 10^{17}$\,GeV and $E_{\mathrm{QG,1}} > 1.7\times 10^{17}$\,GeV for the subluminal and the superluminal scenario, respectively.

\subsection{Applying the Maximum Likelihood Method to the Crab Pulsar with MAGIC}
\label{Subsec:Crab_MAGIC}

The Crab pulsar was also observed and detected by \ac{MAGIC}. The \ac{LIV} analysis performed on the Crab pulsar~\cite{Ahnen:2017wec} focused on the events from the P2 pulse as they reach higher energies, which increases the sensitivity to a \ac{LIV} effect. For this analysis, the authors used $\sim$$326$\,h of excellent quality data. This dataset was analysed with two different methods. Three energy bands (mean energies $\sim$$75$\,GeV, $\sim$$465$\,GeV, and $\sim$$770$\,GeV,) were defined for the analysis but the analysis focused primarily on the two highest. The reason for this choice is that the emission's mechanism is likely to be different between the lowest and the highest energies of the pulse. Therefore the comparison focused on the two high energy bands, which are more likely to arise from the same mechanism that will not affect the search for a \ac{LIV} effect. The first method used was the \ac{PC}, already introduced in \mbox{Section~\ref{Subsec:Crab_VERITAS}}, which yielded the following limits on the \ac{LIV} energy scale: $E_{\mathrm{QG,1}} > 1.1\times 10^{17}$\,GeV and $E_{\mathrm{QG,2}} > 1.4\times 10^{10}$\,GeV for the subluminal scenario, and $E_{\mathrm{QG,1}} > 1.1\times 10^{17}$\,GeV and $E_{\mathrm{QG,2}} > 1.5\times 10^{10}$\,GeV for the superluminal scenario. The second method used by the authors was the \ac{ML} method, here used for the first time to analyse data from a pulsar. The likelihood approach follows what we introduced in Section~\ref{Subsec:MaximumLikelihood}, adapted to the study of events describe by their phase $\phi$ instead of their absolute time $t$. In addition, the likelihood included terms to describe nuisance parameters among which the parameters used to fit the pulse profile and the background events. The former was used to evaluate systematic uncertainties in the analysis, while the latter was particularly important in the case of pulsar located in a Nebula, itself an important and steady source of gamma rays. An extended investigation of the possible origin of systematic uncertainties in this work was performed, including the uncertainty on the absolute energy and flux scale, the possible contribution from events outside the pulse region, and the relatively large uncertainty on the estimation of the distance to the Crab pulsar. In total, the authors estimated the systematic uncertainties on $E_{\mathrm{QG,1}}$ to be less than $42\,\%$ and on $E_{\mathrm{QG,2}}$ to be less than $36\,\%$. 
The obtained limits reached $E_{\mathrm{QG,1}} > 5.5 \times 10^{17}$\,GeV and $E_{\mathrm{QG,2}} > 5.9 \times 10^{10}$\,GeV, including systematic uncertainties, in the subluminal scenario, and $E_{\mathrm{QG,1}} > 4.5 \times 10^{17}$\,GeV and $E_{\mathrm{QG,2}} > 5.3 \times 10^{10}$\,GeV, including systematic uncertainties, in the superluminal scenario. The \ac{ML} method, thus, provided limits a factor 4--5 more stringent than the limits obtained with the \ac{PC} method.

\subsection{Lorentz Invariance Violation Study on a New Vela Pulsar}
\label{Subsec:Vela_HESS}

The Vela pulsar (PSR~J0835-4510), located at $0.29^{+0.08}_{-0.05}$\,kpc~\cite{Caraveo:2001ud} from Earth, was observed by \ac{H.E.S.S.} from March 2013 to April 2015~\cite{Abdalla:2018vik}. In order to reach an energy threshold as low as possible, the analysis only used events recorded by the large 28~m telescope telescope at the centre of the array. The \ac{LIV} analysis~\cite{2015ICRC...34..764C} made use of 24\,h of good quality data from 2013 to 2014. In this period, the telescope recorded about 10,000 pulsed events above $\sim$$20\mathrm{\, GeV}$. The energy range considered for the analysis was 20 to 100\,GeV, yielding a statistics of $\sim$$9300$ excess events associated to the pulsar for a signal to noise ratio of $\sim$$0.025$. The authors used the same \ac{ML} method as described in Section~\ref{Subsec:Crab_MAGIC}. The ON phase region was defined as the interval [0.5,0.6]. The signal template was obtained from the fitting of the low energy ($20\text{--}45$\,GeV) events from the ON phase region by an asymetrical Lorentzian function (for the signal) plus a constant (for the background). This constant is determined from the fitting of events from the OFF phase region chosen as [0.7,1]. 
The authors used dedicated toy Monte Carlo simulations to calibrate their analysis by simulating mock data reproducing Vela's sample characteristics and injecting different simulated phase delays, similar to what was presented in Section~\ref{Subsec:MLonMrk501_MAGIC}. The method exhibits an almost unbiased reconstruction of the \ac{LIV} induced delay. Therefore the results of the distribution of the reconstructed delay, when no \ac{LIV} effect was injected, was used to evaluate the statistical uncertainty of the measurements as well as the systematic uncertainty. Applied to the $45\text{--}100$\,GeV range, the \ac{ML} analysis provided a measurement of the delay $\phi = (-2.0 \pm 5.0_{\mathrm{stat}} \pm 3.0_{\mathrm{sys}}) \times 10^{-2} \, \mathrm{TeV}^{-1}$ compatible with no delay. The results were, therefore, converted to 95\% confidence level lower limits on the linear term $E_{\mathrm{QG,1}}$, yielding $E_{\mathrm{QG,1}} > 4.0\times 10^{15}$\,GeV and $E_{\mathrm{QG,1}} > 3.7\times 10^{15}$\,GeV in the subluminal and superluminal cases, respectively.

\subsection{First Parallel Study of Energy-Dependent Photon Group Velocity and Gamma-ray Absorption on the Same Data Sample}
\label{Subsec:Mrk501_HESS}

As discussed in Section~\ref{Subsec:Mrk501_MAGIC}, Mrk~501 was already observed and studied in the search of \ac{LIV} after a flare detected by \ac{MAGIC} in 2005. In 2014, another flare was detected during a monitoring campaign of \ac{FACT}~\cite{2013JInst...8P6008A}. The alert of this flare triggered observations by the full array of five telescopes of \ac{H.E.S.S.} on the night of 23--24 June 2014. Observations were performed at high zenith angle ($63^{\circ}$ to $65^{\circ}$) leading to a high energy threshold of $\gtrsim$$1$\,TeV. 
The \ac{LIV} analysis on this flare~\cite{Abdalla:2019krx} was done using the \ac{ML} method presented in Section~\ref{Subsec:MaximumLikelihood}. The only noticeable difference was the use of the variable $\eta_n$, defined in Equation~(\ref{eq:eta_n}), as the main likelihood parameter and the explicit mention of the normalization factor depending on $\eta_n$. The sample of events was divided between the 733 events between 1.3\,TeV and 3.25\,TeV, which were used to compute the template and the 662 events above 3.25\,TeV, which were used to compute the likelihood and the best values of $\eta_n$. It is important to note that in this specific analysis, given the high energy threshold of the observations, the low energy template included a potential \ac{LIV} effect. In practice, while the delay in the template is usually taken as null ($D = \eta_n E^n$), here they modelled it as $D = \eta_n (E^n - \overline{E_{\mathrm{T}}}^n)$ where $\overline{E_{\mathrm{T}}}$ is the mean energy of the events in the energy range of the template. As the low energy events were used to built the template, only the 662 high energy events were used in the likelihood in the search for a \ac{LIV} effect in this dataset. The best fitted value of $\eta_n$ were compatible with the Lorentz invariant scenario. 1000 Monte Carlo simulations with no \ac{LIV} effect were used to derive calibrated intervals from which uncertainties were derived. Finally, limits on the energy scale of \ac{LIV} were set to ${E_{\mathrm{QG,1}} > 3.6\times10^{17}\,\mathrm{GeV}}$ 
(${E_{\mathrm{QG,1}} > 2.6\times10^{17}\,\mathrm{GeV}}$) and 
${E_{\mathrm{QG,2}} > 8.5 \times 10^{10}\,\mathrm{GeV}}$ 
(${E_{\mathrm{QG,2}} > 7.3 \times 10^{10}\,\mathrm{GeV}}$)
for the subluminal (superluminal) scenario. These limits include systematic uncertainties, the main one being the determination of the template. Note that, to date, this dataset is the only one that has been used to perform a time of flight study, as described in this section, and a universe transparency study later described in Section~\ref{Subsubsec:HESS_spectral}.

\subsection{First Lorentz Invariance Violation Study on a Gamma-ray Burst Observed with Imaging Atmospheric Cherenkov Telescopes}
\label{Subsec:GRB190114C} 

More that two decades after the proposal by {Amelino-Camelia et al.}, an opportunity presented itself to test the \ac{LIV} on a signal from a \ac{GRB} observed with \acp{IACT}.
The \ac{MAGIC} Collaboration announced a discovery of a \ac{GRB} with \acp{IACT} for the first time ever \cite{2019GCN.23701....1M}\footnote{The discovery of GRB~180720B by \ac{H.E.S.S.} \cite{Arakawa:2019cfc} was announced after the discovery of GRB~190114C.}. A signal from GRB~190114C was detected at energies above 1\,TeV \cite{Acciari:2019dxz}. 
The analysis of this signal for the purpose of testing \ac{LIV} started immediately. The \ac{ML} method was applied (in fact, Equations~(\ref{eq:Likelihood}) and (\ref{eq:PDF}) were adopted from the \ac{LIV} study on GRB~190114C \cite{Acciari:2020kpi}). The most troublesome issue about the analysis was the formulation of the light curve template. The \ac{MAGIC} observations started 62\,s after the burst, almost completely missing the prompt phase, and detecting gamma rays from almost only afterglow phase of the \ac{GRB}. The signal in the TeV energy band was observable until $\sim$$40$\,min after the burst \cite{Acciari:2019dbx}, however it was estimated that only the first 20\,min (the duration of a single observation run) would be relevant for the test of \ac{LIV}. After that, the signal rate became comparable to the background rate, meaning that it would not have considerably improved the sensitivity of the analysis, while at the same time, the systematic effects would have increased.
The \ac{MAGIC} data analysis revealed that during the first 20\,min of observation about 700 gamma rays were detected with the energies in range of $0.3 - \lesssim2$\,TeV. The intrinsic spectral distribution of events was well fitted with a power law \cite{Acciari:2019dbx}. More interestingly, the light curve also demonstrated a monotonic, power law decay of the flux. A monotonic change of flux is no more useful in searches for a spectral dispersion than no change of flux at all would be. A spectral dispersion, applied to a monotonic temporal distribution, would change the rate of a change, but not the functional shape of the distribution. Thus, any effect of a spectral dispersion would be undetectable.
Therefore, in order to perform the \ac{LIV} test, the authors used the light curve model obtained from theoretical inference, and based on the observations performed with the \ac{MAGIC} telescopes and other facilities observing in lower energy ranges \cite{Acciari:2019dbx}. This template was dubbed \textit{theoretical} by the authors of the \ac{LIV} study.
All $\sim$$700$ events were used to calculate the likelihood.
Before estimating the values and the confidence interval of \ac{LIV} parameters, the sensitivity of the method was estimated. This was done by creating 1000 mock data sets and using them to calculate the likelihood. Each mock data set was produced from the original data set by shuffling the arrival times of detected events and then randomly selecting events from this reshuffled data set. In this way, the generated mock data sets consisted of the same number of events as the original data set, with the same energy and temporal distributions. Furthermore, bootstrapping, the procedure of randomly selecting events from the existing (reshuffled) data set, allowed both the energy and temporal distributions to vary in line with their statistical uncertainties. In this way, these uncertainties were propagated to final result. 
Reshuffling, on the other hand, had the role of removing any correlation between the energy and arrival time, if present in the first place. Therefore, if there was any energy-dependent time delay present in the original data set, it would have been washed out by the reshuffling.
After calculating the likelihood for each of the mock data sets, a distribution of the results was made, revealing a bias in the method, for which the final data, obtained on the real data set, were corrected.
The same mock data sets were used to calibrate the confidence interval, as described in Section~\ref{Subsec:MaximumLikelihood}.

Upon correcting for the bias and estimating the confidence interval, the resulting lower bounds on the \ac{LIV} energy scale were as follows: 
${E_{\mathrm{QG,1}} > 5.8\times10^{18}\,\mathrm{GeV}}$ 
(${E_{\mathrm{QG,1}} > 5.5\times}$ ${10^{18}\,\mathrm{GeV}}$) and 
${E_{\mathrm{QG,2}} > 6.3 \times 10^{10}\,\mathrm{GeV}}$ 
(${E_{\mathrm{QG,2}} > 5.6 \times 10^{10}\,\mathrm{GeV}}$)
for the subluminal (superluminal) scenario.

As was already mentioned, the light curve model adopted from \cite{Acciari:2019dbx} was constructed based on observations in lower energies, and theoretical considerations. The power law decay, observed with the \ac{MAGIC} telescopes, in the model is preceded by a rather sharp peak. The peak was before the \ac{MAGIC} observation window, so neither confirmed, nor disproved. So even before obtaining the final results of the \ac{LIV} test, there was a genuine concern that such fast change of the flux was introducing artificially high sensitivity to the \ac{LIV} effects. As a sort of a sanity check, the \ac{LIV} analysis was performed on another light curve template. This template, dubbed ``minimal'', was a step function, with zero value before the burst, and constant value afterwards. Translated to the signal \ac{PDF} (\mbox{Equation~(\ref{eq:PDF})}), it means that there is zero probability of a gamma ray being emitted before the burst, and equal probability of emitting any gamma ray at any time after the burst. This very simple function is clearly not the correct description of the intrinsic light curve. Nevertheless, it avoids sharp peaks not confirmed by observations, consequently, in a sense, minimizing the influence of the light curve template on the sensitivity to the \ac{LIV} effects. This light curve template will cause the likelihood profile to be minimal and flat for small and negative values of the \ac{LIV} parameters, thus preventing the estimation of the bias, and only allowing setting constraints on the subluminal scenario. The results obtained using this minimal model (${E_{\mathrm{QG,1}} > 2.8\times10^{18}\,\mathrm{GeV}}$ and ${E_{\mathrm{QG,2}} > 7.3 \times 10^{10}\,\mathrm{GeV}}$) are compatible with the ones obtained using the theoretical model, meaning that the usage of the theoretical model did not introduce unreasonably high sensitivity into the analysis.

The bounds on the \ac{LIV} energy scale obtained in this study, were comparable to the most constraining lower limits present at that time. 
However, more than confirming the constraints resulting from other studies, the importance of this work was particularly in the fact that it was the first one ever performed on a signal from a \ac{GRB} observed with \acp{IACT}. Especially in the upcoming era of the \ac{CTA}\footnote{\ac{CTA} (\cite{Hassan:2017paq}, \href{https://www.cta-observatory.org/}{https://www.cta-observatory.org/}, accessed 15 July 2021) is an array of Cherenkov telescopes currently being built in two locations. The approved ``Alpha Configuration'' in the Southern Site in Paranal Observatory (Chile) will consist of 14 Medium-Sized Telescopes and 37 Small-Sized Telescopes, covering the area of $\sim$$3$\,km$^2$. The Northern Site will be located in the Roque de los Muchachos Observatory (Spain), consisting of four Large-Sized Telescopes and nine Medium-Sized Telescopes, which will cover the area of $\sim$$0.25$\,km$^2$.}, which carries a promise of observing a few \acp{GRB} each year with significantly larger data samples for every \ac{GRB} \cite{Inoue:2013vy}\footnote{Note that this estimate was obtained for a larger number of telescopes in each site compared to the Alpha Configuration.}, the test of \ac{LIV} on GRB~190114C presents an important stepping stone for the future of \ac{LIV} research.

\subsection{Lorentz Invariance Violation on \textit{Fermi}-LAT Gamma-ray Bursts}
\label{Subsec:Fermi-LAT}
In previous sections we laid out analysis methods and results of different studies performed on the \acp{IACT} data, searching for the signatures of energy dependence in the photon velocity. Results of all these studies are usually compared to the results from a benchmark work by Vasileiou et al. \cite{Vasileiou:2013vra}, where the authors collected four \acp{GRB} observed with the \textit{Fermi}-\ac{LAT} instrument\footnote{Considering our focus on research performed the Vasileiou et al. work is strictly speaking out of the scope of this review. However, it derived some interesting results, and other \ac{LIV} studies are often compared to it, so we will outline its main points.}. 
Vasileiou et al. analysed the \textit{Fermi}-\ac{LAT} data from four bright \acp{GRB} with well determined redshifts: GRB~080916C ($z=4.35 \pm 0.15$ \cite{Greiner:2009pm}), GRB~090510 ($z=0.903 \pm 0.003$ \cite{2009GCN..9352....1O}), GRB~090902B ($z=1.822$ \cite{2009GCN..9873....1C}),
and GRB~090926A ($z=2.1071 \pm 0.0001$ \cite{DElia:2010fwa}). All of these are much farther away than any source used for \ac{LIV} tests with \acp{IACT}. In addition, and unlike the case of GRB~190114C observed with the \ac{MAGIC} telescopes, a quickly variable prompt \ac{GRB} phases were observed in all these four cases. 
An \ac{LIV} test was performed on each of these sources individually, and three different analysis methods were used on each source. 

The \textbf{\acf{PV}}\label{Sec:PairView} method was developed for the purposes of this study. It calculates once the energy-dependent differences in the arrival times for each pair $(i,j)$ of photons in the sample:
 \vspace{6pt}   
\begin{equation}\label{eq:PV}
    l_{\mathrm{i>j}} \equiv \frac{t_{\mathrm{i}} - t_{\mathrm{j}}}{E_{\mathrm{i}}^{{n}}- E_{\mathrm{j}}^{{n}}}.
\end{equation}

The distribution of $l_{\mathrm{i,j}}$ will be peaked at $\eta_{{n}}$ defined in Equation~(\ref{eq:eta_n}), giving the value of the \ac{LIV} parameter.

\textbf{\Acf{SMM}}\label{Sec:SMM} \cite{Vasileiou:2013vra, deAlmeida:2012cx} is analogue to the \ac{ECF} method, which was previously applied to the \ac{MAGIC} sample of Mrk~501 and explained in Section~\ref{Subsec:Mrk501_MAGIC}. It employs the aforementioned fact that an application of a spectral dispersion to a data set will decreases sharpness of the light curve. While the \ac{ECF} method maximizes the power in the selected time interval, the \ac{SMM} measures the sharpness of the light curve, e.g.,
    \vspace{6pt}
\begin{equation}\label{eq:SMM}
    S(\eta_{n}) = \sum_{i=1}^{N-\rho}\log\left(\frac{\rho}{t'_{i+\rho} - t'_{i}} \right),
\end{equation}
after applying an opposite dispersion as described in Equation~(\ref{eq:t_prime}). $\rho$ is a fixed parameter making sure that events which are very close together are not considered in the denominator, because that would dominate the function. The intrinsic light curve is expected to be the sharpest one. Therefore, $\eta_{{n}}$ for which the light curve is the sharpest, will be the measure of the spectral dispersion present in the data sample. 

\textls[-12]{The third and final method used was the \ac{ML}, which we already described in \mbox{Section~\ref{Subsec:MaximumLikelihood}}}. 
Final limits on the \ac{LIV} energy scale were obtained for each source individually, by taking average value of results obtained from three different methods and after accounting for systematic effects (see Table 5 in \cite{Vasileiou:2013vra}).
The most constraining lower limits resulted from the GRB~090510: $E_{\mathrm{QG,1}} > 2.2\times10^{19}$\,GeV, $E_{\mathrm{QG,2}} > 4.0\times10^{10}$\,GeV for the subluminal, and $E_{\mathrm{QG,1}} > 3.9\times10^{19}$\,GeV, $E_{\mathrm{QG,1}} > 3.0\times10^{10}$\,GeV for the superluminal scenarios.
The staggering lower limits on the linear term, surpassing the Planck energy are the reason why every other \ac{LIV} study is compared to this one. 
Interestingly, GRB~090510 was the one with the smallest redshift in the sample, and the only one with $z<1$. However, it is also the only one in the sample which was classified as a short \ac{GRB}, while the other three were long \acp{GRB}, with emission spread over somewhat longer time. Moreover, the highest energies in all four data sets were detected from GRB~090510. 
While the authors at first considered combining the results from all four sources into one single bound on the \ac{LIV} energy scale, they gave up on the idea because the result from GRB~090510 was so much more constraining than the other three \acp{GRB} that a combination would not significantly increase the lower limit.

It should be noted that the \textit{Fermi}-\ac{LAT} detector is sensitive in the energy range of 20\,MeV--300\,GeV. The higher energy part of this band partially overlaps with the \acp{IACT} sensitivity range ($\sim$$30$\,GeV---few tens of TeV). 
However, in the overlapping energy region, the \textit{Fermi}-\ac{LAT} sensitivity deteriorates with increasing energy, while the opposite is true for \acp{IACT}\footnote{For a comparison of sensitivities of various current and future instruments see \cite{CTAperformance} and references therein.}. 
Lower energy gamma rays, detectable with \textit{Fermi}-\ac{LAT}, are not absorbed by the \ac{EBL}, thus enabling \textit{Fermi}-\ac{LAT} to detect sources at significantly higher redshifts than \acp{IACT}, increasing the sensitivity to \ac{LIV} effects. On the other hand, \textit{Fermi}-\ac{LAT} reaches significantly lower energies than \acp{IACT}, which limits its sensitivity to \ac{LIV} effects. These characteristics will be important when we compare different results in Section~\ref{Sec:ResultsComparison}.

\section{Modified Photon Interactions}
\label{Sec:Modiefied_photon_interactions}

Very soon after the modified photon dispersion relation was introduced, it has been realised that it can have consequences on kinematics and dynamics of the processes (see e.g., \cite{Kifune:1999ex,Aloisio:2000cm,2004NJPh....6..188A}). In some quantum electrodynamics processes, modifications of dispersion relation may cause the change of the reaction energy threshold. On the other hand, some processes forbidden by energy-momentum conservation law in Lorentz invariant scenario, may become allowed if the Lorentz symmetry is broken. In this chapter, we will look more closely into several of these phenomena. 

\subsection{Testing Lorentz Invariance Violation with Universe Transparency}
\label{Subsec:UniverseTransparency}

The universe is filled with low energy photon fields such as extragalactic background light (\ac{EBL}), \ac{CMB} and \ac{RB}. Gamma rays traversing cosmological distances scatter off those photons creating electron-positron pairs. Consequently, their flux, observed from Earth, is attenuated \cite{Gould_Schreder:1967, Stecker:1992wi, Biteau:2015xpa, Abdalla:2018sxi}.~The \ac{EBL} is responsible for the attenuation of gamma rays in $10\text{--}10^{5}$\,GeV range, which roughly corresponds to observable energy range of current \acp{IACT}. Unfortunately, direct \ac{EBL} measurements are obstructed by bright foreground emissions, mainly zodiacal light \cite{Hauser:2001xs}, which makes it hard to determine its precise spectrum (for more information about the \ac{EBL}, photon-photon interactions and the opacity of the universe to gamma rays we referee the reader to \cite{Franceschini:2021wkr} and references therein). To tackle this problem, different phenomenological approaches predicting overall \ac{EBL} spectrum have been followed.
Remarkably, \ac{EBL} models obtained through different methodologies, such as Franceschini et al. \cite{Franceschini:2008tp}, \mbox{Dom\'{i}nguez et al. \cite{Dominguez:2010bv}} and Gilmore et al. \cite{Gilmore:2011ks}, are in a good agreement. These models were tested on \ac{VHE} data from sets of \ac{AGN} by current \acp{IACT} \cite{H.E.S.S.:2017odt,Acciari:2019zgl,Abeysekara:2019ybp}. Those tests were done presuming Lorentz invariance.

The gamma-ray spectrum observed from Earth is usually written as a convolution of the source intrinsic spectrum and the \ac{EBL} attenuation effect:

\begin{equation}\label{eq:obs_spectrum}
    \Phi_{\mathrm{obs}}(E) = \Phi_{\mathrm{int}}(E(1+z_{\mathrm{s}})) \times \mathrm{e}^{-\tau(E,z_{s})},
\end{equation}
where $E$ is the observed gamma-ray energy and $z_{\mathrm{s}}$ is the redshift of the observed source.
$\tau(E,z_{\mathrm{s}})$ is the optical depth, dependent on the two aforementioned parameters and is given~by\footnote{Henceforward, when denoted with prime physical quantity is written in the comoving frame at which the interaction occurs. When prime does not occur, it is written in the observer's frame of reference.}:

\begin{equation}\label{eq:optical_depth}
\tau(E,z_{\mathrm{s}})=\int_{0}^{z_{\mathrm{s}}}\frac{dl}{dz} d z \int_{-1}^{1} \frac{1-\cos\theta^{\prime}}{2}  d\cos\theta^{\prime}\int_{\epsilon_{th}^{\prime}}^{\infty} \sigma_{\gamma \gamma}\left(s\right) n\left(\epsilon^{\prime}, z\right) d \epsilon^{\prime}.
\end{equation}

In this expression,
\begin{itemize}
    \item $\epsilon^{\prime}$ denotes the energy of an \ac{EBL} photon in the comoving frame, while $n\left(\epsilon^{\prime}, z\right)$ is the comoving number density of \ac{EBL} photons per unit energy.
    \item The probability of the interaction between a gamma ray and background photons is given by the cross section $\sigma_{\gamma \gamma}(s)$, where $s$ is the square of the center of mass energy. In the gamma-ray energy range relevant for \acp{IACT}, by far the most dominant channel is the Breit--Wheeler process of electron-positron pair creation \cite{Breit_Wheeler1934PhRv}. 
    \item The angle of interaction between a gamma ray and \ac{EBL} photons is indicated by $\theta^{\prime}$.
    \item $\epsilon_{\mathrm{th}}^{\prime}$ denotes the \ac{EBL} energy reaction threshold for electron-positron pair creation, i.e., the minimal energy of an \ac{EBL} photon, in the comoving frame, necessary for the reaction to take place. Derived from the kinematics laws of special relativity, it can be expressed as:
\begin{align}
\label{eq:energy_threshold}
   \epsilon_{\mathrm{th}}^{\prime}=\frac{2m_{\mathrm{e}}^{2}c^{4}}{E^{\prime}(1-\cos\theta^{\prime})}. 
\end{align}
    The threshold energy, and its changes due to modifications of the special relativity kinematics, will play a vital role in constraining $E_{\mathrm{QG}}$. 
    \item The final integral accounts for the distance traveled by the gamma ray, assuming flat $\Lambda\mathrm{CDM}$ cosmology:
\begin{align}
\label{eq:ComovingDistance}
    \frac{dl}{dz} = \frac{c}{H_{0}\left(1+z\right) \sqrt{\Omega_{\mathrm{m}}\left(1+z\right)^{3}+\Omega_{\Lambda}}}
\end{align}
\end{itemize}

Beyond the gamma-ray horizon $(\tau(E,z_{\mathrm{s}})=1)$ the universe becomes progressively opaque for \ac{VHE} gamma rays (for further readings on this topic in connection with \acp{IACT} we suggest \cite{Blanch_Martinez_1:2005APh} and references therein). For a sources at redshift 0.034, which is a redshift of Mrk\,501, gamma-ray horizon is around 10 TeV \cite{Franceschini:2008tp}. 
When doing calculations, one must be careful to take into account cosmic expansion and notice that measurements are affected by a factor $(1+z)$. Namely, $E$ and $\epsilon$ change along the line of sight inversely proportional to $(1+z)$; for example, $\epsilon = \epsilon^{\prime}/(1+z)$.

\subsubsection{Influence of Lorentz Invariance Violation on Universe Transparency}
\label{Subsubsec:when_it_all_began}

Detected gamma-ray emission up to $\sim$$22$\,TeV from Mrk\,501 \cite{Aharonian:1999vy} in 1997 by \ac{HEGRA} experiment\footnote{The HEGRA experiment was a system of five \acp{IACT} decommissioned in 2002 \cite{Puhlhofer:2003ir}.} hinted that the universe is more transparent to \ac{VHE} gamma rays than expected. One possible solution to this newly arisen problem was the aforementioned modification of photon dispersion relation. Added terms in the photon dispersion relation can cause a change in the energy threshold for pair creation, consequently leading to changes in the gamma-ray absorption. In this scenario, the new energy reaction threshold is \cite{2003APh....19..245B}:

\begin{equation}\label{eq:modified_energy_threshold}
\epsilon_{\mathrm{th}}^{\prime}= \frac{2 m_{\mathrm{e}}^{2} c^{4}}{E^{\prime} (1-\cos\theta^{\prime})} - \frac{S}{2 (1-\cos\theta^{\prime})}\left(\frac{E^{\prime}}{E_{\mathrm{QG,}n}}\right)^{n}E^{\prime}
\end{equation}

\begin{figure}
    \includegraphics[width=0.7\textwidth]{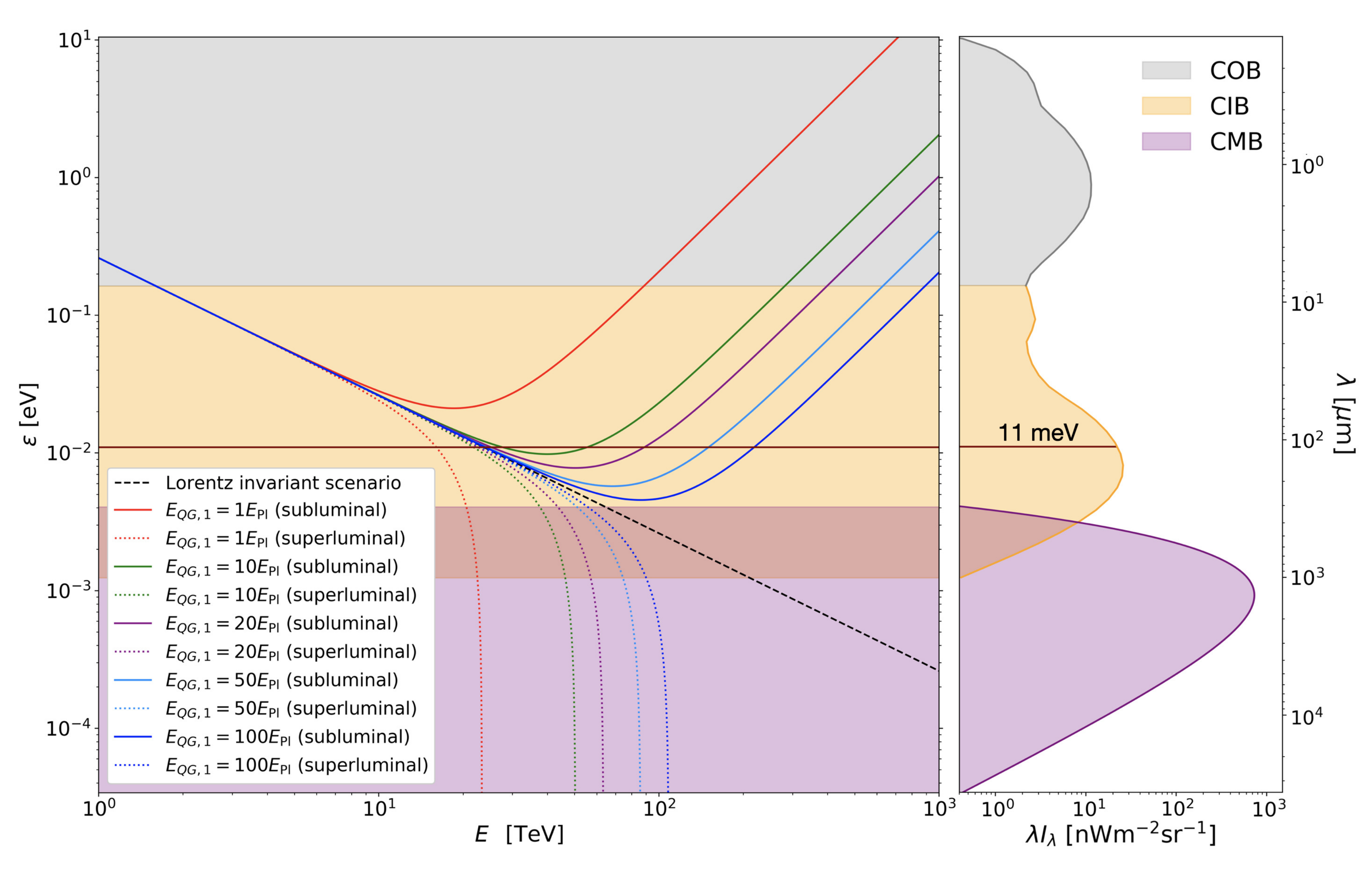}
    \caption{\label{Fig:Energy_Threshold_SED} ({\textbf{Left}}) Energy of the background photons at threshold $(\epsilon_{\mathrm{th}})$ for the pair-production reaction as a function of a gamma-ray energy $(E)$. The black dashed line represents the Lorentz invariant scenario, while solid and doted lines represent \ac{LIV} subluminal and superluminal scenarios, respectively. Five different values of $E_{\mathrm{EG,1}}$ were considered. ({\textbf{Right}}) Spectral energy distributions of the \ac{CMB} and the two constituents of the \ac{EBL} (cosmic optical background and cosmic infrared background) were produced using the \ac{EBL} model by Dom\'{i}nguez et al. \cite{Dominguez:2010bv}.
    }
\end{figure}

Changes in energy reaction threshold are depicted in Figure~\ref{Fig:Energy_Threshold_SED} for a head-on collision and $z = 0$. As defined in Section~\ref{Subsec:ModifiedDispRel}, $S=+1$ for superluminal, and $S=-1$ for subluminal behaviour. This modified energy reaction threshold has been derived under two assumptions: (i) the standard energy-momentum conservation law is maintained in \ac{LIV} scenario, and (ii) \ac{LIV} affects only dispersion relation for photons, while electrons remain unaffected. 
Indeed, the effects of \ac{LIV} on electrons were strongly constrained by independent studies\footnote{Modifications of the electron dispersion relation were tested on the 100\,MeV synchrotron radiation from the Crab nebula in \cite{Jacobson:2002ye}. The lower limit on $E_{\mathrm{QG,1}}$ for electrons was set to at least seven orders of magnitude above $E_{\mathrm{Pl}}$.}.
Therefore, the majority of studies considering electromagnetic interaction rely on the assumption that only the photon dispersion relation is modified by~\ac{LIV}.

Modifications in the energy reaction threshold could lead to changes in the observed spectra of a distant source, depending on the \ac{LIV} scale \cite{Kifune:1999ex, Protheroe_Meyer:2000PhLB, AmelinoCamelia:2000zs, Jacob:2008gj, Biteau:2015xpa, Tavecchio:2015rfa}.
In the superluminal behaviour $(S=+1)$, modifications in the photon dispersion relation will cause lowering of the energy reaction threshold for the electron-positron pair creation. In that case, gamma rays would be absorbed by lower energy photon fields than in the Lorentz invariant case. For example, a 50\,TeV gamma ray in Lorentz invariant scenario does not have enough energy to reach the reaction threshold with \ac{CMB} photons. However, in a \ac{LIV} superluminal scenario, for sufficiently low values of $E_{\mathrm{QG}}$, the reaction threshold will be reached (see Figure~\ref{Fig:Energy_Threshold_SED}). This would lead to additional depletion of the most energetic photons resulting in a steeper observed spectrum. Still, so far no way was found to unambiguously disentangle the effects caused by the lowering of the photon-photon energy threshold due to \ac{LIV}, from the effects arising due to Lorentz invariant \ac{EBL} attenuation, or intrinsic properties of the source such as a spectral cut off. This is arguably the main reason why all experimentally set limits on $E_{\mathrm{QG}}$ using the universe transparency to gamma rays were derived for the subluminal behaviour only. 
In the subluminal scenario $(S=-1)$, modifications of the photon dispersion relation will lead to an increase of the energy reaction threshold, resulting in a reduced opacity of the universe to \ac{VHE} gamma rays. 
Moreover, the reaction threshold as a function of the gamma-ray energy will have a global minimum \cite{Jacob:2008gj} as can be seen in Figure~\ref{Fig:Energy_Threshold_SED}. 
Note that there is no equivalent minimum in the Lorentz invariant nor \ac{LIV} superluminal scenario, since $\epsilon_{\mathrm{th}}^{\prime}$, as defined in Equation~(\ref{eq:energy_threshold}), is a monotonous function of the gamma ray energy. Contrary to the Lorentz invariant scenario and the \ac{LIV} superluminal scenario, for which once the reaction energy threshold is reached the reaction is allowed for all gamma rays with energies above the reaction energy threshold, a pair creation in the \ac{LIV} subluminal scenario is kinematically forbidden for gamma-ray energies higher than the reaction energy threshold. 
The existence of the global minimum implies that the energy domain of \ac{EBL} photons, as targets for absorption of \ac{VHE} gamma rays, would be reduced, regardless of the gamma-ray energy. Consequently, a certain number of gamma rays would evade absorption and thus reach the Earth. This most particularly holds for gamma rays with energies above the position of the energy threshold minimum.

\begin{figure}
     \includegraphics[width=0.7\textwidth]{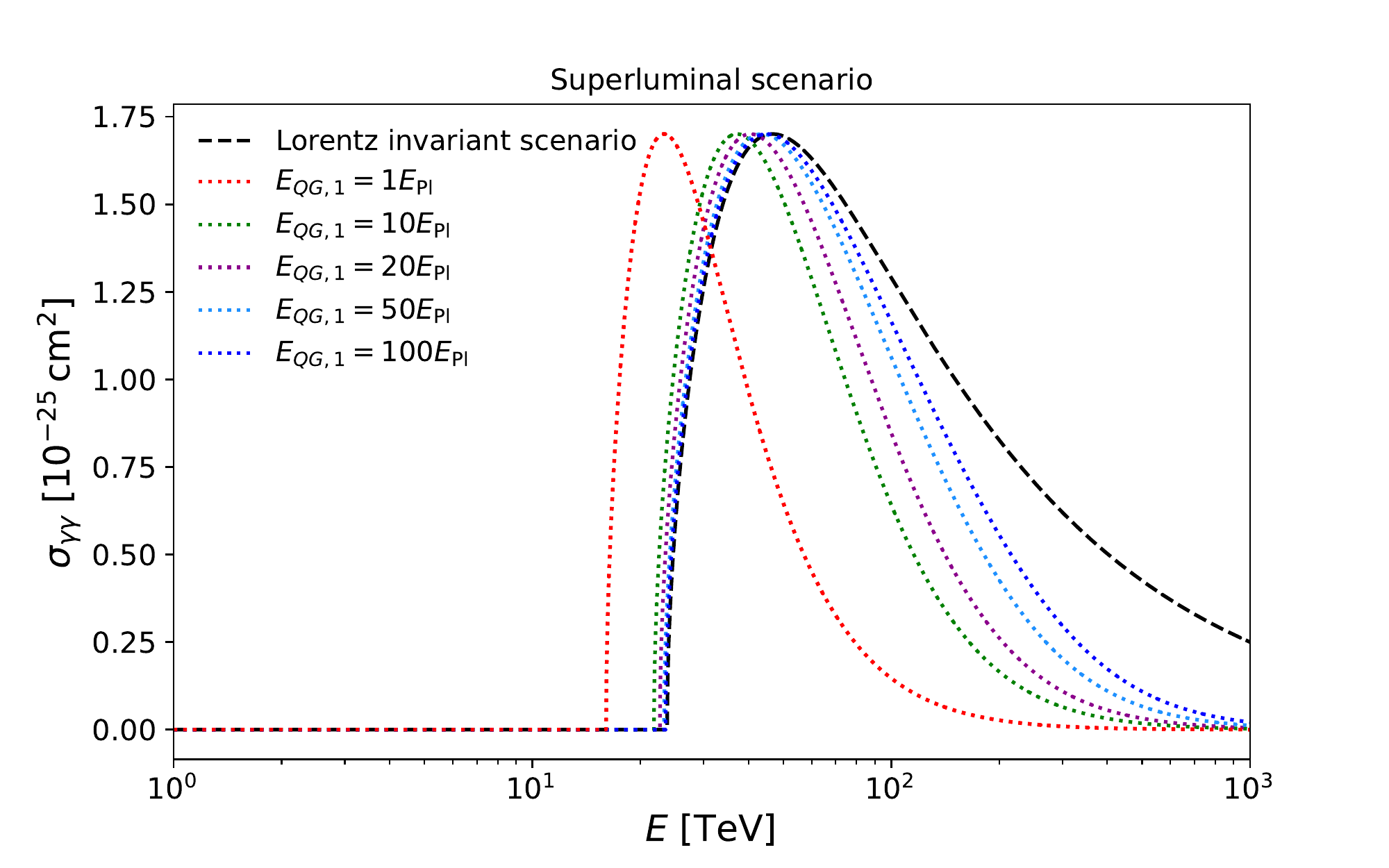}
     \includegraphics[width=0.7\textwidth]{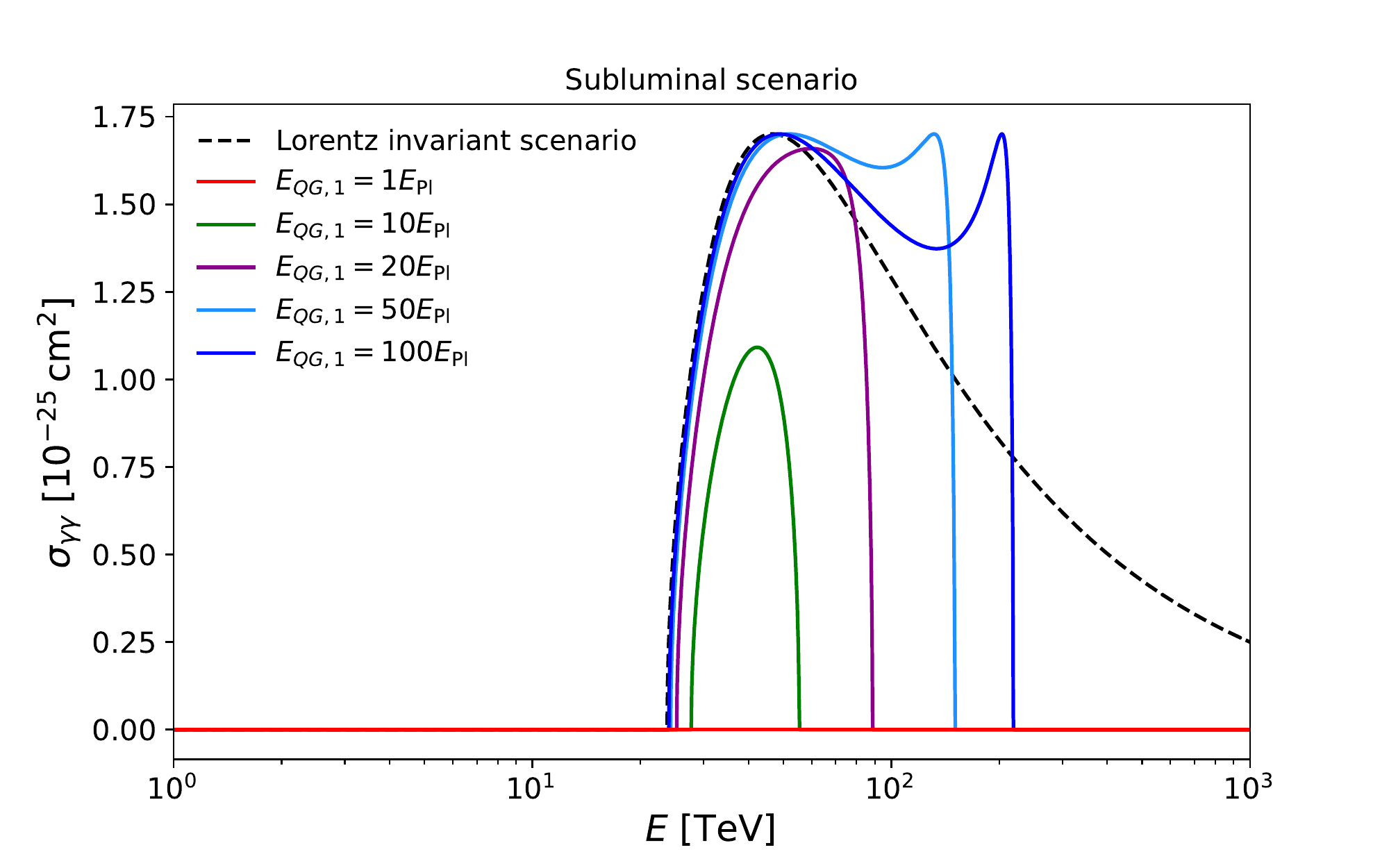}
     \caption{\label{Fig:crossection2} The Breit--Wheeler cross section as a function
 of the gamma-ray energy for the background photon energy of 11\,meV and a head-on collision. The same five values of $E_{\mathrm{EG,1}}$ as in Figure~\ref{Fig:Energy_Threshold_SED} were considered. (\textbf{Top}) The black dashed line represents the Lorentz invariant scenario while doted lines represent \ac{LIV} superluminal scenario. (\textbf{Bottom}) The black dashed line represents the Lorentz invariant scenario while full lines represent \ac{LIV} subluminal scenario.} 
\end{figure}

The aforementioned Breit--Wheeler cross section, as a function of the gamma-ray energy, is shown in Figure~\ref{Fig:crossection2}. In the Lorentz invariant scenario, it is represented with a black dashed line. Once the reaction energy threshold is reached, the cross section rises quickly. At the gamma-ray energy roughly twice the threshold energy \cite{1996ApJ...456..124M}, the Lorentz invariant cross section reaches its maximal value of $ \simeq$$1.70 \times 10^{-25}~\mathrm{cm}^{2}$~\cite{DeAngelis:2013jna}. Afterwards, as the gamma-ray energy increases, the cross section drops and asymptotically approaches zero. 
In the superluminal scenario, the energy reaction threshold is lower than in the Lorentz invariant scenario. Moreover, lower $E_{\mathrm{QG,}1}$ results in lower reaction threshold. The cross section shape remains the same as in the Lorentz invariant scenario, although, it becomes narrower as $E_{\mathrm{QG,}1}$ decreases and reaches its maximum at lower gamma-ray energies.
A somewhat more interesting development of the Breit--Wheeler cross section occurs in a subluminal \ac{LIV} scenario. There are three distinct cases: (i) As we saw in Figure~\ref{Fig:Energy_Threshold_SED}, for $E_{\mathrm{QG,}1}$ low enough, the reaction energy threshold will never be reached. Consequently, the cross section will be zero for all gamma-ray energies (red full line in the bottom panel of Figure~\ref{Fig:crossection2}). (ii) For higher values of $E_{\mathrm{QG,}1}$, the horizontal line will be crossed twice. Hence, there will be a lower and an upper reaction energy thresholds, and the reaction will be possible for gamma-ray energies between these thresholds. For relatively low $E_{\mathrm{QG,}1}$, this interval will be narrow, and the cross section will never reach its maximum possible value of $ \simeq$$1.70 \times 10^{-25}~\mathrm{cm}^{2}$ (green and violet full lines in the bottom panel of Figure~\ref{Fig:crossection2}). (iii)~For even higher values of $E_{\mathrm{QG,}1}$, the gamma-ray energy interval between the reaction energy thresholds will be wide enough for the cross section to reach its maximum possible value. Moreover, the cross section will start to decrease with increasing gamma-ray energy, roughly following the shape of the Lorentz invariant cross section.
However, as the gamma-ray energy rises, the cross section reaches a local minimum, and starts increasing to reach its maximum possible value once again, just below the upper reaction threshold. Once the threshold is reached, the cross section is cut off (light and dark blue full lines in the bottom panel of Figure~\ref{Fig:crossection2}). If one continued to increase the value of $E_{\mathrm{QG,}1}$, the second peak would become sharper and move towards higher energies, and the intermediate part of the cross section would more closely follow the Lorentz invariant cross section.
In addition to these three cases, there are borderline cases. Between cases (i) and (ii), for $E_{\mathrm{QG,}1}$ precisely such that the horizontal line in Figure~\ref{Fig:Energy_Threshold_SED} is a tangent to the energy threshold line, the reaction energy threshold will be reached at precisely one gamma-ray energy, and the cross section will be a vertical line at that energy, and zero elsewhere. Between cases (ii) and (iii), the cross section would reach its maximum possible value, and monotonically drop to zero.

\begin{figure}
    \includegraphics[width=0.6\textwidth]{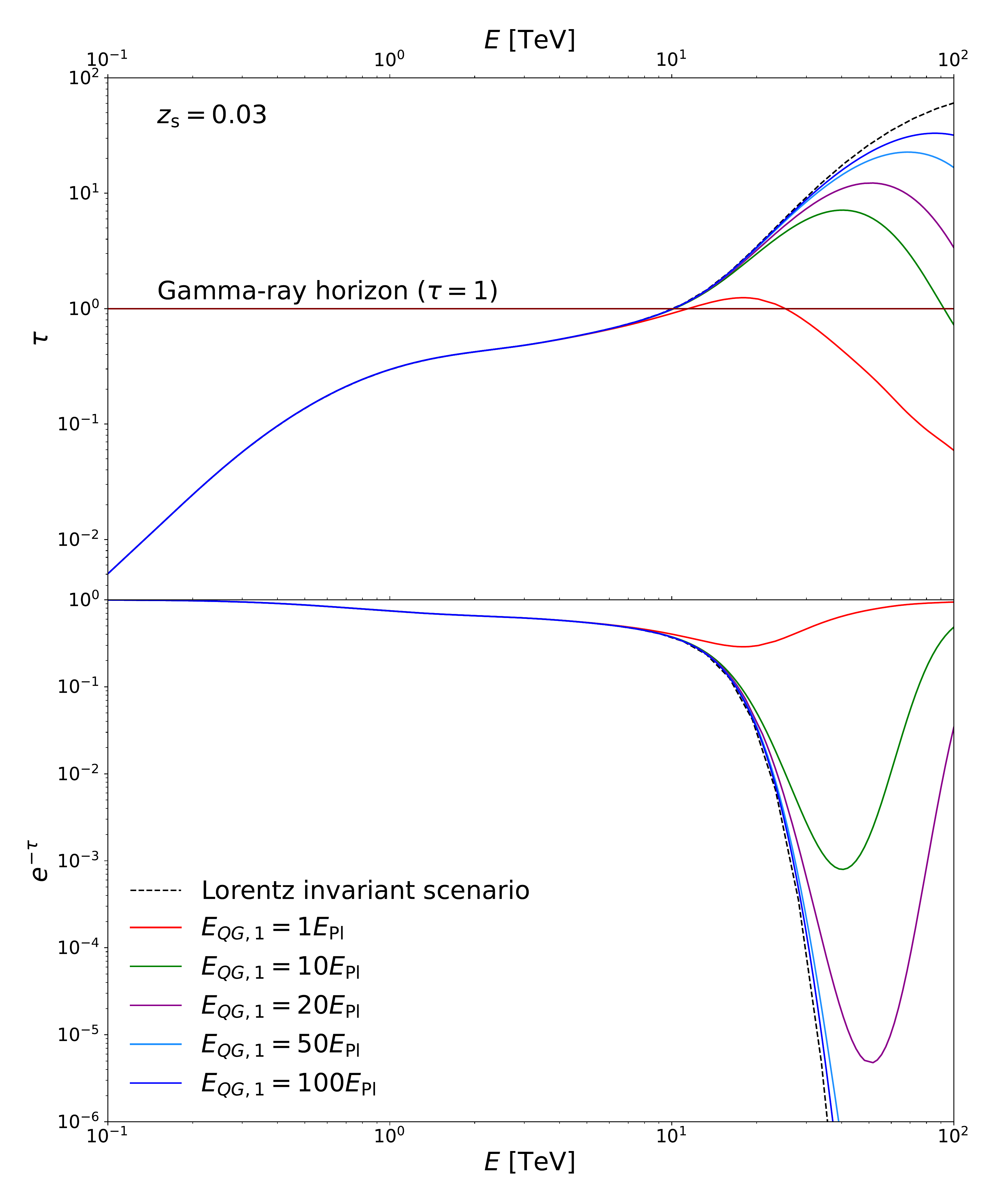}
    \caption{\label{Fig:Exp(-tau)_Tau_together} (\textbf{Top}) Optical depth $(\tau)$ as a function of a gamma-ray energy $(E)$ for a hypothetical source at $z_{\mathrm{s}} = 0.03$. The black dashed line represents the Lorentz invariant scenario, while solid lines represent \ac{LIV} subluminal scenario. Five different values of $E_{\mathrm{EG,1}}$ were considered. Gamma-ray horizon is denoted with the maroon line. (\textbf{Bottom}) The absorption coefficient $(e^{-\tau})$ as a function of a gamma-ray energy $(E)$.}
\end{figure}

After seeing how modifications of photon dispersion relation influence the reaction energy threshold and the cross section, now it is time to see how the optical depth changes due to those modifications, since all experimental limits on $E_{\mathrm{QG}}$, based on the universe transparency, were set only for the subluminal scenario, we will focus only on this scenario. In Figure~\ref{Fig:Exp(-tau)_Tau_together} we depicted a hypothetical gamma-ray absorption for a source at redshift $z_{\mathrm{s}} = 0.03$ and gamma-ray energies up to 100\,TeV\footnote{Gamma-ray energy range up to 100\,TeV was chosen to be compatible with previously published results based on universe transparency to gamma rays. To best of our knowledge there is no publication in which the optical depth behaviour beyond this limit has been investigated. For this reason we departed from the usual energy range used in other figures in this section.}, assuming different values of $E_{\mathrm{QG,1}}$. In the top panel of Figure~\ref{Fig:Exp(-tau)_Tau_together} the gamma-ray horizon is denoted with a maroon line. As previously mentioned, beyond the gamma-ray horizon the universe becomes increasingly opaque for \ac{VHE} gamma rays and thus the probability of their detection is diminishing. In the Lorentz invariant scenario, once the gamma-ray horizon is reached, the optical depth only increases. On the other hand, in the subluminal \ac{LIV} scenario, the optical depth has a global maximum, different for different energies depending on the value of $E_{\mathrm{QG}}$, after which it decreases again. At some point it goes below the gamma-ray horizon allowing the gamma rays to evade absorption, which would lead to the recovery of the photon flux. The higher the $E_{\mathrm{QG}}$ scale, the higher the energy of the gamma-ray at which the recovery would occur. The absorption coefficient $(e^{-\tau})$, as a function of a gamma-ray energy $(E)$ is depicted in the bottom panel of Figure~\ref{Fig:Exp(-tau)_Tau_together} and shows how the survival probability of the photons behaves in this scenario.

It should be noted that there are other phenomena other than \ac{LIV} which could leave imprints in the spectra of observed sources. Most notable are the axion-like particles, into which \ac{VHE} gamma rays can oscillate in the presence of the external magnetic field. Nevertheless, imprints which axion-like particles and \ac{LIV} would potentially leave could be mutually distinguished. Namely, axion-like particle imprints should be independent of the source redshift, while at the same time dependent on the magnetic field. The opposite holds for the \ac{LIV}. For now, these two effects are being investigated separately, even in studies investigating both of them (see, e.g., \cite{Biteau:2015xpa}). For more information about axion-like particles and their searches with \acp{IACT}, we refer the interested reader to  \cite{2021Univ....7..185B, CTA:2020hii}.

\subsubsection{Testing Lorentz Invariance Violation on Universe Transparency}
\label{Subsubsec:Biteau_Williams}

The first experimental test of \ac{LIV} on the \ac{EBL} absorption of gamma rays using data from \acp{IACT} was performed by Biteau $\&$ Williams \cite{Biteau:2015xpa}. The authors derived a simplified expression for optical depth using analytical methods. Furthermore, they constructed the \ac{EBL} spectrum using 86 published gamma-ray spectra of 30 blazars with well-established redshifts. A total of $\sim$270,000 gamma rays constituted this gamma-ray sample. The \ac{EBL} model was described through eight free parameters, denoted with $A$. In the \ac{LIV} scenario, $E_{\mathrm{QG}}$ was added as one additional free parameter to the analysis, using the optical depth dependency on \ac{LIV} described in the previous section. Furthermore, the \ac{EBL} parameters were allowed to vary with $E_{\mathrm{QG}}$. In order to investigate the possible effects of \ac{LIV}, Biteau $\&$ Williams compared spectra of the aforementioned blazars under the assumption of Lorentz invariance on the one hand, and under the assumption of \ac{LIV} on the other hand. The effect of \ac{LIV} was quantified using a \ac{TS} defined as follows, with $\mathcal{L}=\exp (-TS/2)$:

\begin{equation}\label{eq:TS_Biteau_Williams}
TS=\chi^{2}\left(E_{\mathrm{QG}}, A_{\mathrm{QG}}\right)-\chi^{2}\left(\infty, A_{\infty}\right).
\end{equation}

The term $\chi^{2}\left(E_{\mathrm{QG}}, A_{\mathrm{QG}}\right)$ represented the best fit in the \ac{LIV} scenario. The $\chi^{2}\left(\infty, A_{\infty}\right)$ represented the best fit in the Lorentz invariant scenario since, as mentioned in Section~\ref{Subsec:ModifiedDispRel}, letting $E_{\mathrm{QG,1}} \rightarrow \infty$ leads to the Lorentz invariant photon dispersion relation. 

Biteau $\&$ Williams adopted the formalism of \cite{Jacob:2008gj}, which presumes that \ac{LIV} affects both photons and leptons equally. Under that assumption, the last term in the energy threshold expression (Equation~(\ref{eq:modified_energy_threshold})) gains another factor of $(1-2^{-n})$. One of the 86 spectra selected for this study was the spectrum of Mrk\,501 historical flare detected by \ac{HEGRA} in 1997 \cite{Aharonian:1999vy}. 
When performing the analysis on the originally published Mrk\,501 spectrum, Biteau $\&$ Williams showed that $E_{\mathrm{QG,1}}$ was approximately $E_{\mathrm{Pl}}$ at $4\sigma$ level.
However, when a newly derived spectrum of the same data set (from \cite{Aharonian:2000xr}) was used, the significance decreased to $2.4\,\sigma$. The re-analysed spectrum had better energy resolution, therefore it excluded the initially obtained highest energy data point. This example demonstrates how a single spectrum, and the most energetic photons in it, can greatly influence the final result. The first experimentally set $95\%$ confidence level lower limit, using universe transparency to \ac{VHE} gamma rays, was found to be $E_{\mathrm{QG,1}}>9.5 \times 10^{18}$\,GeV. This value changed to $E_{\mathrm{QG,1}}>8.6 \times 10^{18}$\,GeV when a $10\%$ systematic uncertainty on the energy scale (typical for \acp{IACT}) was accounted for.

\subsubsection{The Most Constraining Limits Based on Single Source Analysis}
\label{Subsubsec:HESS_spectral}

Seventeen years after the historical flare from Mrk\,501, which triggered discussions about the influence of \ac{LIV} to universe's transparency to \ac{VHE} gamma rays, another bright flare brought Mrk\,501 in the spotlight once again \cite{Cologna:2016cws}.
This time it was used to constrain $E_{\mathrm{QG}}$ by the \ac{H.E.S.S.} Collaboration using two independent channels \cite{Abdalla:2019krx}. The test of energy-dependent photon group velocity was described in Section~\ref{Subsec:Mrk501_HESS}, while the spectral analysis will be discussed here. Only the possible subluminal behaviour for linear or quadratic contributions was investigated. The pair production cross-section was calculated according to \cite{Fairbairn:2014kda}. In this approach, the modified expression for the square of the center of mass energy $s$ can be written as:

\begin{equation}\label{eq:modified_center_of_mass}
s= 2 E^{\prime} \epsilon^{\prime} (1-\cos\theta^{\prime})+ S \left(\frac{E^{\prime}}{E_{\mathrm{QG,}n}}\right)^{n} E^{\prime 2}
\end{equation}

The dependence of the cross section on $s$ is considered to be the same in \ac{LIV} and Lorentz invariant scenarios. The cross section as a function of a gamma-ray energy for the background photon of 11\,meV and a head-on collision is depicted in Figure~\ref{Fig:crossection2}.

To quantify the possible effects of \ac{LIV} on a spectra of this bright flare, authors defined a \ac{TS} similar to Equation~(\ref{eq:TS_Biteau_Williams}):

\begin{equation}\label{eq:TS_HESS}
\mathrm{TS}=\chi^{2}\left(E_{\mathrm{QG}}\right)-\chi^{2}\left(E_{\mathrm{QG}} \rightarrow \infty\right).
\end{equation}

However, unlike the \ac{TS} defined in Equation~(\ref{eq:TS_Biteau_Williams}), this \ac{TS} did not contain free-varying \ac{EBL} parameters. The \ac{EBL} model of Franceschini et al. \cite{Franceschini:2008tp} was used and the optical depth was calculated in a standard way, as described in Equation~(\ref{eq:optical_depth}). The $\chi^{2}$ values in Equation~(\ref{eq:TS_HESS}) were obtained by varying $E_{\mathrm{QG}}$ logarithmically and fitting the measured spectrum of a flare by an assumed intrinsic spectrum convoluted with the \ac{EBL} attenuation effect. The intrinsic spectrum was assumed to be a simple power law. From the \ac{TS} profiles, lower limits on $\mathrm{E}_{\mathrm{QG}}$ at $95\%$ confidence level were set to $E_{\mathrm{QG},1}>2.6 \times 10^{19}$\,GeV and $E_{\mathrm{QG},2}>7.8 \times 10^{11}$\,GeV for the linear and quadratic contributions, respectively.

A peculiarity of this work lies in the fact that the same data set was used to put the constraints on $E_{\mathrm{QG}}$ via energy-dependent time delay and the universe transparency effects, which we will discuss in more detail in Section~\ref{Sec:ResultsComparison}.

\subsubsection{On How the Most Constraining Limits Were Obtained}
\label{Subsubsec:Lang_et_al}

At the moment of writing this review, the best lower limits on $E_{\mathrm{QG}}$ obtained by testing the universe transparency were obtained by Lang et al. \cite{Lang:2018yog}. The starting data sample consisted of 111 spectra from 38 different sources available in the online catalogue TeVCat\footnote{\url{http://tevcat.uchicago.edu/}.}, of which only a subset was used for setting constraints on \ac{LIV}. In order to select only the relevant sources for probing \ac{LIV}, the authors defined the attenuation $a(E, z)$ as the ratio between the measured $J_{\text {meas}}(E)$ and the intrinsic $J_{\text {int}}(E, z)$ spectra of each source:

\begin{equation}\label{Lang_attenuation}
a(E, z)=e^{-\tau(E, z)}=\frac{J_{\text {meas}}(E)}{J_{\text {int}}(E, z)}
\end{equation} 

Lang et al. calculated the ratio between the attenuation assuming \ac{LIV} and the attenuation in the Lorentz invariant scenario, at the maximum energy $(E_{\mathrm{max}})$ measured in a given spectrum. Only 18 spectra from 6 different sources for which the $a_{LIV}/a_{LI}$ ratio differed by at least $10\%$ and which could be used to further constrain $E_{\mathrm{QG}}$ were selected for further~analysis.

In general, the intrinsic spectrum is obtained via the process of so-called deabsorption which consists in reverting the Equation~(\ref{eq:obs_spectrum}). In previous studies, deabsorption was done under the assumption of Lorentz invariance, not taking into account \ac{LIV} in this step of the analysis. In order to rectify it, Lang et al. used an energy interval which they call fiducial region. It is defined as the energy range starting at the lowest measured spectral point. The highest spectral point is the last one at which the difference between the fluxes assuming \ac{LIV} and assuming Lorentz invariance are indistinguishable, considering the measurement uncertainties. Therefore only measured spectral points from bins that satisfy the following condition were used to determine the intrinsic spectrum:

\begin{equation} \label{Lang_fiducal_bins}
\frac{a_{\mathrm{LIV}}}{a_{\mathrm{LI}}} \leq \frac{J_{\text {meas}}(E)+\rho \sigma\left(J_{\text {meas }}(E)\right)}{J_{\text {meas}}(E)}
\end{equation}

Throughout their work, the authors assumed $\rho = 1$, implying the tightest energy interval and hence leading to the most conservative limits on $E_{\mathrm{QG}}$. Every intrinsic spectrum $(J_{\mathrm{int}})$ was modeled as a power law with an exponential cut off. For each selected spectrum $(J_{\mathrm{int}})$, the energy spectrum on Earth $(J_{\mathrm{cal}})$ was computed for multiple $E_{\mathrm{QG,}n}$ values using
\vspace{6pt}
\begin{equation}\label{lang_J_cal}
J_{\mathrm{cal}}=a_{\mathrm{LIV}} \times J_{\mathrm{int}} 
\end{equation}

Subsequently, all computed spectra were compared with the complete measured spectra $J_{\mathrm{meas}}$ using a log-likelihood method. Finally, the authors combined the likelihood results from all the sources to achieve the best possible sensitivity.

In their work, Lang et al. report $2\sigma$ confidence level lower limits on $E_{\mathrm{QG}}$ obtained with three different \ac{EBL} models, using the same procedure. The most conservative limits were derived using the \ac{EBL} model by Dom\'{i}nguez et al. \cite{Dominguez:2010bv} and were set to be $E_{\mathrm{QG,1}}>6.85 \times 10^{19}$\,GeV and $E_{\mathrm{QG,2}}>1.56 \times 10^{12}$\,GeV.

\subsection{Constraints on Violation of Lorentz Invariance from Atmospheric Showers Initiated by Multi-TeV Photons}
\label{Subsec:LIVonShowers}

\textls[-15]{The imaging technique of Cherenkov telescopes relies on recording flashes of Cherenkov} light produced in the atmosphere by ultrarelativistic particles constituting extensive air showers. When a \ac{VHE} gamma ray enters the Earth's atmosphere, it is absorbed in the Coulomb field of an atomic nucleus in the air, creating an electron--positron pair. Each created particle carries approximately one half of the primary gamma ray's energy. Leptons are emitting additional gamma rays through \textit{bremsstrahlung}, each of which again go through the process of pair production. In that way, an electromagnetic cascade is created. The pair-creation is, fundamentally, the same process as the gamma-gamma interaction in which \ac{VHE} gamma rays get absorbed by the \ac{EBL}. Therefore, if the gamma-gamma interaction was affected by modifying the photon dispersion relation, it would also influence the development of particle showers in the atmosphere. This interesting notion was proposed by {Rubtsov, Satunin \& Sibiryakov} in \cite{Rubtsov:2013wwa} and tested on data from \ac{HEGRA} and \ac{H.E.S.S.} in~\cite{Rubtsov:2016bea}. 

A shower development is governed by the \ac{B-H} process. In particular, the depth of the first interaction in the atmosphere is exponentially distributed with the mean value inversely proportional to the cross section \cite{Bethe:1934za}

\begin{equation}\label{eq:BHprocess}
    \sigma_{\mathrm{BH}} = \frac{28Z^2 \alpha^3 \hbar^2 c^2}{9m_\mathrm{e}^2 c^4} \left( \log\frac{183}{Z^{1/3}} - \frac{1}{42} \right).
\end{equation}

Here $Z$ is the atomic number of the nucleus, $\alpha$ is the fine structure constant, and $m_\mathrm{e}$ is the electron mass. 
In \ac{LIV} scenario, the cross section will not change significantly for superluminal photons, unless the threshold for photon decay is reached. However, in that case, the photon decay will be the dominant process, making the \ac{LIV} influence on the \ac{B-H} process negligible. On the other hand, if photons are subluminal, the \ac{B-H} cross section becomes strongly suppressed, leading to the suppression factor

\begin{equation}\label{eq:BHprocessLIV}
    \frac{\sigma_{\mathrm{BH}}^{\mathrm{LIV}}}{\sigma_{\mathrm{BH}}} \simeq \frac{12m_\mathrm{e}^2 c^4 E_{\mathrm{QG,2}}^2}{7 E_{\gamma}^4}  \log\frac{E_{\gamma}^4}{2m_\mathrm{e}^2 c^4 E_{\mathrm{QG,2}}^2}
\end{equation}

As a consequence, the shower development in the \ac{LIV} scenario will be impeded. The first gamma-gamma interaction will occur deeper in the atmosphere, and the effect will be more pronounced for higher gamma-ray energies. This will lead to showers reaching their maximal sizes also deeper in the atmosphere. Height of the shower maximum is an important parameter in \acp{IACT} data analysis (see, e.g., \cite{deNaurois:2009ud}). Depending on the experimental setup and the details of the data analysis, changes in the \ac{B-H} cross section might lead to the showers induced by the most energetic gamma rays being misrepresented and excluded from further analysis. Ultimately, this will result in an apparent cut off in the spectrum at the high end. 

{Rubtsov, Satunin \& Sibiryakov} applied this method on two independent measurements of the Crab nebula spectrum. The first one was obtained by the \ac{HEGRA} Collaboration, based on 385 h of observations performed between 1997 and 2002 \cite{Aharonian:2004gb}. The highest energy bin in the spectrum was centered at 75\,GeV. 
The analysis method used to compare the spectra in Lorentz invariant versus \ac{LIV} scenarios, and to determine limits to the \ac{LIV} energy scale was based on \ac{ML} method similar to the one used in \cite{Abdalla:2019krx}. 
They set a limit to the \ac{LIV} energy scale to $E_{\mathrm{QG,2}} > 2.1\times 10^{11}$\,GeV. 
The second sample was the Crab nebula spectrum measured by the \ac{H.E.S.S.} Collaboration, based on 4.4 h during the flaring episode in March 2013 \cite{Abramowski:2013qea}. In this case, the spectrum was determine up to $\sim$$40$\,TeV. Because of the smaller data set and lower energies reached, the result was less constraining: $E_{\mathrm{QG,2}} > 1.3\times 10^{11}$\,GeV.

Note that this effect would be opposite to the one caused by modified absorption of gamma rays on the \ac{EBL} (described in Section~\ref{Subsec:UniverseTransparency}). The constraints on the \ac{LIV} energy scale set in the work by {Rubtsov, Satunin \& Sibiryakov} were lower than the ones obtained based on the universe transparency to gamma rays. However, the constraints based on the universe transparency were obtained assuming that shower development, and consequently measurement with \acp{IACT}, is not modified by \ac{LIV}. The work by {Rubtsov, Satunin \& Sibiryakov} tested and validated that assumption.

\subsection{Constraints on Lorentz Invariance Violation Based on Photon Stability}
\label{Subsec:PhotonDecay}
A modifying term in the photon dispersion relation can be treated as the mass term in the (unmodified) dispersion relation of a massive particle. Assigning a photon a mass in the superluminal scenario renders it unstable and prone to decay. A superluminal photon of energy $E_{\gamma}$ can:
\begin{itemize}
    \item decay into an electron--positron pair
    
    \begin{equation*}
        \gamma \longrightarrow e^{+} + e^{-}.
    \end{equation*}
    
    This reaction becomes possible under condition \cite{Martinez-Huerta:2016azo}
    
    \begin{equation}\label{eq:photonDecay}
        E_{\mathrm{QG,}n} > E_{\gamma} \left(\frac{E_{\gamma}^2}{4m_{\mathrm{e}}^2 c^4} -1\right)^{1/n},
    \end{equation} 
    where $m_{\mathrm{e}}$ is the electron mass. \\
    \item split into multiple photons
    \begin{equation*}
        \gamma \longrightarrow N\gamma,
    \end{equation*}
    with the dominant channel being splitting into three photons \cite{Astapov:2019xmt}.
\end{itemize}

While the photon splitting has no reaction threshold, and is kinematically allowed for every superluminal photon, the process rate (see \cite{Astapov:2019xmt}) is significantly smaller than the photon decay rate \cite{Martinez-Huerta:2016azo, Albert:2019nnn}. 

Both processes have a similar effect on the observed spectra from astronomical sources, i.e., spectra will be attenuated at higher energies. Since there is no reaction threshold for the photon splitting, the attenuation will happen gradually. The photon decay rate quickly increases with the gamma-ray energy once the reaction threshold is reached. Consequently, it will be manifested as a cut off in the spectrum.
The effects are similar to spectral attenuation due to gamma-ray absorption on the \ac{EBL}. Therefore, in order to test for photon decay or photon splitting, one needs to exclude the possibility of \ac{EBL} absorption. An obvious choice of a \ac{VHE} gamma-ray source for these studies is the Crab nebula. Its spectrum reaches energies well above 100\,TeV \cite{Amenomori2019:Tibet, Abeysekara:2019edl}, while at the same time, because of its small distance from the Earth, \ac{EBL} absorption is virtually negligible.
These effects were independently tested for on Crab nebula spectral measurements in several studies (see, \mbox{e.g., \cite{Martinez-Huerta:2016azo, Astapov:2019xmt, Satunin:2019gsl}}), however, the results setting substantially stronger constraints came not from any \ac{IACT} experiment, but from the \ac{HAWC} Collaboration\footnote{As the name says, \ac{HAWC} (\href{https://www.hawc-observatory.org/}{https://www.hawc-observatory.org/}, accessed 15 July 2021) is a water Cherenkov experiment located in Parque Nacional Pico de Orizaba in~Mexico.} \cite{Albert:2019nnn}.
The photon decay was used to constrain both liner and quadratic terms, obtaining 
$E_{\mathrm{QG,1}} > 2.2\times 10^{22}$\,GeV and
$E_{\mathrm{QG,2}} > 0.8\times 10^{14}$\,GeV, respectively. 
The photon splitting was used only for the quadratic term, resulting in a much stronger constraint than the photon decay, $E_{\mathrm{QG,2}} > 1.0\times 10^{15}$\,GeV. 
Sheer moments before concluding this review, a very exciting result was published by \ac{LHAASO}\footnote{\ac{LHAASO} (\href{http://english.ihep.cas.cn/lhaaso/}{http://english.ihep.cas.cn/lhaaso/}, accessed 15 July 2021) is a hybrid detector of cosmic and gamma rays located in Daocheng, Sichuan province, China \cite{Zhen:2019lnn}.}, announcing a detection of gamma rays with energies up to 1.4\,PeV from 12~sources \cite{Cao2021}. 
Based on these measurements the \ac{LHAASO} Collaboration performed a similar study as \ac{HAWC}. They searched for a cut of in spectra of the two sources with the highest energies LHAASO~J0534+2202 (Crab nebula) and LHAASO~J2032+4102 (the source which the 1.4\,PeV event was associated with) \cite{LHAASO:2021opi}. Due to significantly higher spectral measurements by \ac{LHAASO}, the resulting constraints on the \ac{LIV} energy scale were also higher. Specifically, their most constraining limits were based on the analysis of LHAASO~J2032+4102 spectrum. After including the systematic uncertainties, the limits were set to $E_{\mathrm{QG,1}} > 1.2\times 10^{24}$\,GeV and $E_{\mathrm{QG,2}} > 1.1\times 10^{15}$\,GeV, when the photon decay was considered. As in the \ac{HAWC} analysis, only quadratic term was constrained based on photon splitting, resulting in $E_{\mathrm{QG,2}} > 2.0\times 10^{16}$\,GeV. 

The scope of this review are studies performed with \acp{IACT}. Nevertheless, we included these results from other type of observatories for comparison purposes. In addition, similar studies could be performed with \acp{IACT}. With the prospect of the \ac{CTA} to be commissioned in the next few years, and a recent result from the \ac{MAGIC} Collaboration measuring the Crab nebula spectrum up to 100\,TeV \cite{Acciari:2020lmm}, feasibility of a similar study with \acp{IACT} does not seem so far-fetched.

\section{Summary and Discussion}
\label{Sec:ResultsComparison}

In this section we will review and compare the results presented in previous sections. The results are summarised and listed chronologically in Table~\ref{Tab:Results}. A quick glance already reveals that the constraints, both on linear and quadratic terms, are the strongest in the case of the photon stability measurements \cite{Albert:2019nnn} by \ac{LHAASO}, surpassing the Planck energy by four orders of magnitude in the case of the linear term. Apparently, that is the effect the most sensitive to \ac{LIV}. However, as already stated in Section~\ref{Subsec:PhotonDecay}, photon decay and photon splitting are processes only allowed in superluminal scenario, and cannot be used to constrain $E_{\mathrm{QG}}$ in the subluminal scenario. 
Nevertheless, it was important to constrain this \ac{LIV} effect. Possible photon decay and photon splitting competes with modified absorption of gamma rays on EBL. Without a confirmation of the photon stability to these substantially high values of $E_{\mathrm{QG}}$, it would be virtually impossible to resolve and independently search for \ac{LIV} effects of modified universe transparency to gamma rays. As it happens, the second most constraining bounds were set precisely on measurements of the universe transparency for gamma rays, or gamma-gamma interaction. The most notable result was obtained by Lang et al.~\cite{Lang:2018yog} through a simultaneous analysis of spectra of several \acp{AGN}. As already argued, combining different sources, should wash out dependence on the properties of a given source and, therefore, help to lift the degeneracy between the \ac{LIV} and source intrinsic effects.
The main characteristics of a desirable source for this method are large source distance and the highest spectral energy measurement. Combining various sources in a single study enables the most pronounced characteristic of each source to be fully exploited. However, the method heavily relies on the \ac{EBL} modeling and the assumptions made on the intrinsic source spectra. Uncertainties of \ac{EBL} models, as well as discrepancies between the different models, are the predominant source of systematic effects. Lang et al. considered this uncertainty by presenting the results of their analysis using three different \ac{EBL} models. In this review, we presented the most conservative result.

\startlandscape
\begin{specialtable}
\widetable
\caption{\label{Tab:Results}Chronological census of bounds on the \ac{LIV} energy scale as reported in the respective publications. The columns show from left to right: (i) \ac{LIV} effect used to probe the \ac{QG}, (ii)~name of the observed source (when more than one source was used, we give the numner of the sources in the sample), (iii) type of the observed source, (iv) source redshift for extragalactic sources, and distance in kpc for sources in Milky Way, (v) analysis method used to perform the test, (vi) lower limit on the linear term in the modified photon dispersion relation, (vii) lower limit on the quadratic term in the modified photon dispersion relation, (viii) the experiment which produced the data sample, (ix) reference for the respective publication, (x) section in which the result was discussed in this work. Markers (+) and ($-$) in columns (vi) and (vii) represent superluminal and subluminal behaviours, respectively. The lower limits are expressed on the 95\,\% confidence level. Symbol $^{\dagger}$ represents the best fit result in case where the estimated parameter was $>2\sigma$ away from Lorentz invariant scenario.}
\footnotesize
\begin{tabular}{llclclllll}

\toprule
\textbf{Effect}	& \textbf{Source} & \textbf{Type} & \textbf{Distance} & \textbf{Method} & $\bm{E_{\mathrm{QG,1}}}$\,\textbf{[GeV]} & $\bm{E_{\mathrm{QG,2}}}$\,\textbf{[GeV]} & \textbf{Instr.} & \textbf{Ref.} & \textbf{Sec.}\\

\midrule
Time delay & Mrk~421 & \ac{AGN} & $z = 0.031$ & band comparison & $(-)\quad0.4 \times 10^{17}$ &  & Whipple & \cite{1999PhRvL..83.2108B} & \ref{Subsec:Mrk421_Whipple}\\

\midrule
Time delay & Mrk~501 & \ac{AGN} & $z = 0.034$ & \ac{ECF} & $(-)\quad2.1 \times 10^{17}$ & $(-)\quad2.6 \times 10^{10}$ & \ac{MAGIC} & \cite{2008PhLB..668..253M} & \ref{Subsec:Mrk501_MAGIC}\\

\midrule
Time delay & Mrk~501 & \ac{AGN} & $z = 0.034$ & \ac{ML} & $(-)\quad3.0\times{10^{17}}^{~\dagger}$ & $(-)\quad5.7\times{10^{10}}~^{\dagger}$ & \ac{MAGIC} & \cite{2009APh....31..226M} & \ref{Subsec:MLonMrk501_MAGIC}\\

\midrule
\multirow{2}{*}{Time delay} & \multirow{2}{*}{PKS~2155-304\quad\quad\quad\quad} & \multirow{2}{*}{\ac{AGN}} & \multirow{2}{*}{$z = 0.116$} & \ac{MCCF} & $(-)\quad7.2\times10^{17}$ & $(-)\quad1.4\times10^{9}$ & \multirow{2}{*}{\ac{H.E.S.S.}} & \multirow{2}{*}{\cite{Aharonian:2008kz}} & \multirow{2}{*}{\ref{Subsec:PKS2155_HESS}}\\
& & & & \ac{CWT} & $(-)\quad5.2\times10^{17}$ &  & & &\\

\midrule
Time delay & PKS~2155-304 & \ac{AGN} & $z = 0.116$ & \ac{ML} & $(-)\quad2.1\times10^{18}$ & $(-)\quad6.4\times10^{10}$ & \ac{H.E.S.S.} & \cite{2011APh....34..738H} & \ref{Subsec:PKS2155_HESS}\\

\midrule
\multirow{2}{*}{Time delay} & \multirow{2}{*}{GRB~090510} & \multirow{2}{*}{\ac{GRB}} & \multirow{2}{*}{$z = 0.9$} & \ac{PV}, \ac{SMM}, & $(-)\quad2.2\times10^{19}$ & $(-)\quad4.0\times10^{10}$ & \multirow{2}{*}{\ac{LAT}} & \multirow{2}{*}{\cite{Vasileiou:2013vra}} & \multirow{2}{*}{\ref{Subsec:Fermi-LAT}}\\
& & & & and \ac{ML} & $(+)\quad3.9\times10^{19}$ & $(+)\quad3.0\times10^{10}$ & & &\\

\midrule
\multirow{3}{*}{Time delay} & \multirow{3}{*}{Crab} & \multirow{3}{*}{Pulsar} & \multirow{3}{*}{$d=2$\,kpc} & \multirow{1}{*}{\ac{PC}} & $(-) \quad 3.0\times 10^{17}$ & $(-) \quad 7.0\times 10^{9}$ & \multirow{3}{*}{\ac{VERITAS}} & \multirow{1}{*}{\cite{2011ICRC....7..256O}} & \multirow{3}{*}{\ref{Subsec:Crab_VERITAS}}\\
& & & & \multirow{2}{*}{\ac{DisCan}} & $(-) \quad 1.9\times 10^{17}$ &  & & \multirow{2}{*}{\cite{2013ICRC...33.2768Z}} & \\
& & & & & $(+) \quad 1.7\times 10^{17}$ &  & & &\\

\midrule
\multirow{2}{*}{Time delay} & \multirow{2}{*}{PG~1553+113} & \multirow{2}{*}{\ac{AGN}} & \multirow{2}{*}{$z = 0.49$} & \multirow{2}{*}{\ac{ML}} & $(-)\quad4.1\times10^{17}$ & $(-)\quad2.1\times10^{10}$ & \multirow{2}{*}{\ac{H.E.S.S.}} & \multirow{2}{*}{\cite{Abramowski:2015ixa}} & \multirow{2}{*}{\ref{Subsec:PG1553_HESS}}\\
& & & & & $(+)\quad2.8\times10^{17}$ & $(+)\quad1.7\times10^{10}$ & & &\\

\midrule
\multirow{2}{*}{Time delay} & \multirow{2}{*}{Vela} & \multirow{2}{*}{Pulsar} & \multirow{2}{*}{$d=0.3$\,kpc} & \multirow{2}{*}{\ac{ML}} & $(-) \quad 4.0\times 10^{15}$ &  & \multirow{2}{*}{\ac{H.E.S.S.}} & \multirow{2}{*}{\cite{2015ICRC...34..764C}} & \multirow{2}{*}{\ref{Subsec:Vela_HESS}}\\
& & & & & $(+) \quad 3.7\times 10^{15}$ &  & & &\\

\midrule
Universe transparency & Multiple (30) & \ac{AGN} & $z = 0.019-0.287$ & \ac{TS} & $(-)\quad8.6\times10^{18}$ &   & Multiple & \cite{Biteau:2015xpa} & \ref{Subsubsec:Biteau_Williams}\\

\midrule
\multirow{2}{*}{Bethe--Heitler} & \multirow{2}{*}{Crab} & \multirow{2}{*}{Nebula} & \multirow{2}{*}{$d = 2$\,kpc} & \multirow{2}{*}{\ac{ML}} &  & $(-)\quad2.1\times10^{11}$ & \ac{HEGRA} & \multirow{2}{*}{\cite{Rubtsov:2016bea}} & \multirow{2}{*}{\ref{Subsec:LIVonShowers}}\\
& & & & & & $(-)\quad1.3\times10^{11}$ & \ac{H.E.S.S.} & &\\

\midrule
\multirow{4}{*}{Time delay} & \multirow{4}{*}{Crab} & \multirow{4}{*}{Pulsar} & \multirow{4}{*}{$d=2$\,kpc} & \multirow{2}{*}{\ac{PC}} & $(-) \quad 1.1 \times 10^{17}$ & $(-) \quad 1.4 \times 10^{10}$ & \multirow{4}{*}{\ac{MAGIC}} & \multirow{4}{*}{\cite{Ahnen:2017wec}} & \multirow{4}{*}{\ref{Subsec:Crab_MAGIC}}\\
& & & & & $(+) \quad 1.1 \times 10^{17}$ & $(+) \quad 1.5 \times 10^{10}$ & & &\\
& & & & \multirow{2}{*}{\ac{ML}} & $(-) \quad 5.5 \times 10^{17}$ & $(-) \quad 5.9 \times 10^{10}$ & & &\\
& & & & & $(+) \quad 4.5 \times 10^{17}$ & $(+) \quad 5.3 \times 10^{10}$ & & &\\

\midrule
\multirow{2}{*}{Time delay} & \multirow{3}{*}{Mrk~501} & \multirow{3}{*}{\ac{AGN}} & \multirow{3}{*}{$z = 0.034$} & \multirow{2}{*}{\ac{ML}} & $(-) \quad 3.6\times10^{17}$ & $(-) \quad 8.5 \times 10^{10}$ & \multirow{3}{*}{\ac{H.E.S.S.}} & \multirow{3}{*}{\cite{Abdalla:2019krx}} & \multirow{2}{*}{\ref{Subsec:Mrk501_HESS}}\\
& & & & & $(+) \quad 2.6\times10^{17}$ & $(+) \quad 7.3 \times 10^{10}$ & & &\\
Universe transparency & & & & \ac{TS} & $(-)\quad2.6\times10^{19}$ & $(-)\quad7.8 \times 10^{11}$ & & & \ref{Subsubsec:HESS_spectral}\\

\midrule
Universe transparency\quad\quad\quad\quad\quad & Multiple (6) & \ac{AGN} & $z = 0.031-0.188 $ & \ac{TS} & $(-)\quad6.9\times10^{19}$ & $(-)\quad1.6 \times 10^{12}$ & Multiple & \cite{Lang:2018yog} & \ref{Subsubsec:Lang_et_al}\\

\midrule
Photon decay & \multirow{2}{*}{Multiple (4)} & \multirow{2}{*}{Galactic} & \multirow{2}{*}{$d = 1.55 - 2.37$\,kpc} & \multirow{2}{*}{\ac{TS}} & $(+)\quad2.2\times10^{22}$ & $(+)\quad0.8\times10^{14}$ & \multirow{2}{*}{\ac{HAWC}} & \multirow{2}{*}{\cite{Albert:2019nnn}} & \multirow{2}{*}{\ref{Subsec:PhotonDecay}}\\
Photon splitting & & & & & & $(+)\quad1.0\times10^{15}$ & & &\\

\midrule
\multirow{2}{*}{Time delay} & \multirow{2}{*}{GRB~190114C} & \multirow{2}{*}{\ac{GRB}} & \multirow{2}{*}{$z = 0.4245$} & \multirow{2}{*}{\ac{ML}} & $(-)\quad5.8\times10^{18}$ & $(-)\quad6.3\times10^{10}$ & \multirow{2}{*}{\ac{MAGIC}} & \multirow{2}{*}{\cite{Acciari:2020kpi}} & \multirow{2}{*}{\ref{Subsec:GRB190114C}}\\
 & & & & &  $(+)\quad5.5\times10^{18}$ & $(+)\quad5.6\times10^{10}$ & & \\

\midrule
Photon decay & \multirow{2}{*}{J2032+4102} & Stellar & \multirow{2}{*}{$d = 1.4$\,kpc} & \multirow{2}{*}{\ac{TS}} & $(+)\quad1.2\times10^{24}$ & $(+)\quad1.1\times10^{15}$ & \multirow{2}{*}{\ac{LHAASO}} & \multirow{2}{*}{\cite{LHAASO:2021opi}} & \multirow{2}{*}{\ref{Subsec:PhotonDecay}}\\
Photon splitting & & cluster & & & & $(+)\quad2.0\times10^{16}$ & & &\\

\bottomrule
\end{tabular}
\end{specialtable}
\finishlandscape

A second look at the Table~\ref{Tab:Results} clearly shows that analyses based on photon interactions result in stronger constraints on $E_{\mathrm{QG}}$ compared to the ones based on photon time of flight.
A study by the \ac{H.E.S.S.} Collaboration was the only one so far in which both tests were performed on the same data set \cite{Abdalla:2019krx}. Granted, the two analyses were performed independently, the time of flight test ignoring modified gamma-ray absorption and vice versa. Nevertheless, it allowed a more direct comparison of the two effects of \ac{LIV}. The constraints based on absorption of gamma rays on the \ac{EBL} were more stringent by one order of magnitude on the quadratic term, and two orders of magnitude on the linear term, compared to the constraints based on energy-dependent time delay. 
Apart from the source distance and the detected energy, when it comes to time delay studies, another important property comes into play. Indeed, fast variability of flux is crucial in constraining emission times of individual photons (see Table~\ref{Tab:TOFsensitivity}). 
It should be noted that, while changes in flux do not strongly interfere with analysis based on \ac{EBL} absorption, it is extremely important that the spectrum remains constant. A change in the spectrum would introduce additional uncertainties and hence lower the analysis sensitivity.
Therefore, one may argue that a faster flux variability is needed to increase the sensitivity of time of flight tests. However, even the most sensitive of the time of flight analyses (\cite{Vasileiou:2013vra} for the linear, and \cite{Abdalla:2019krx} for the quadratic term) are below the sensitivities of analyses based on photon interactions, 
suggesting that (assuming the same $E_{\mathrm{QG}}$) \ac{LIV} affects interactions more strongly than it does the photon group velocity.
So, is there a point in testing energy-dependent photon group velocity, when we were able to set much stronger bounds on $E_{\mathrm{QG}}$ through tests of modified photon interactions? As we pointed out in Section~\ref{Subsec:ModifiedDispRel}, the theory of \ac{QG} has not been formulated yet. Consequently, we do not know what the effects of \ac{QG} are. It is quite possible that the photon group velocity is affected by the \ac{LIV}, while interactions remain unaffected, or the other way around. Another possibility is that both interactions and propagation are affected, but on different scales, effectively introducing separate and different values of $E_{\mathrm{QG}}$. It should be remembered that the modified dispersion relation (Equation~(\ref{eq:moddispastro})) is merely a mathematical model facilitating experimental tests of \ac{LIV}, and, in the most general case, different values of $E_{\mathrm{QG}}$ are applicable in different cases. Yet another imaginable scenario is that both subluminal and superluminal behaviours occur as different effects of \ac{QG}. In that case, they might start to manifest at different energy scales, resulting in an even more complex expression for the photon group velocity (Equation~(\ref{eq:photonvelocity})).

Considering the time delay studies alone, we observe a gradual increase of the lower limits on $E_{\mathrm{QG}}$. Of course, a study declaring a stricter limit (or a detection for that matter) is more likely to be published, however we would like to argue that this improvement is a result of several circumstances: (i) improvement on the performances of detectors allowed observations of sources at larger redshifts with better sampling, (ii) longer operations increased the probability of observing transient phenomena as flaring \acp{AGN} and \acp{GRB}, as well as increased statistics on observations of pulsars, and (iii) analysis techniques, the \ac{ML} in particular, have been refined over time, allowing for higher analysis sensitivity. 
A notable exception is the result obtained on the observation of GRB~090510 with \textit{Fermi}-\ac{LAT}. Published in 2013, the study by Vasileiou et al.~\cite{Vasileiou:2013vra} still holds the record for the most stringent bound on $E_{\mathrm{QG,1}}$. Let us analyse where this sensitivity came from. 
As conveyed in Table~\ref{Tab:TOFsensitivity}, the sensitivity of time of flight analyses increases with the energy of the gamma rays within the sample ($E_{\mathrm{max}}$) and the redshift of the source ($z_{\mathrm{s}}$), and decreases with the timescale of flux variations ($t_{\mathrm{var}}$). GRB~090510 has the second and third criterion well satisfied. The data set used in the analysis was taken from a time interval of less than 5\,s (see Figure~1 in~\cite{Vasileiou:2013vra}), significantly shorter than any other sample covered in this review. At the same time, the redshift of $\sim$$0.9$ is more than two times larger than the second furthest source considered in Table~\ref{Tab:Results}. Apparently, the downside of the low highest energy in the sample (31\,GeV) is more than well compensated by these two advantageous characteristics. In the case of the quadratic term, the relationship between $E_{\mathrm{max}}$, $t_{\mathrm{var}}$, and $z_{\mathrm{s}}$ is somewhat different. $E_{\mathrm{QG}}$ is still inversely proportional to the variability timescale, however $t_{\mathrm{var}}$ now enters through a square root, decreasing its influence on the analysis sensitivity. Furthermore, $E_{\mathrm{QG}}$ depends on the source redshift as $z_{\mathrm{s}}^{\sim2/3}$ (in contrast to $z_{\mathrm{s}}^{\sim1}$ in case of the linear term). Given that all sources are at redshifts smaller than 1, the influence of redshift is weaker on the quadratic term. 
Weaker influences of variability and redshift make more room for the influence of $E_{\mathrm{max}}$. Now, the tables have turned. The bound on $E_{\mathrm{QG,1}}$ set using \ac{MAGIC} observation of GRB~190114C~\cite{Acciari:2020kpi} was only about a factor 4 (7) below the limit from GRB~090510 for the subluminal (superluminal) behaviour. In the quadratic term, the influence of the highest gamma-ray energy is more pronounced than in the linear term. So, the constraint on $E_{\mathrm{QG,2}}$ became more stringent because of $\lesssim$$2$\,TeV photons in the GRB~190114C sample. 
Actually, regarding the quadratic case, the strongest constraint came from the \ac{H.E.S.S.} analysis of \ac{AGN} Mrk~501, due to $\sim$$20$\,TeV photons in the sample \cite{Abdalla:2019krx}.

Emission time from pulsars is excellently constrained, with the variability timescales down to $t_{\mathrm{var}}$$\sim$$10$\,ms. Furthermore, gamma rays from pulses are detected up to $\sim$$7$\,TeV. Unfortunately, their relatively close proximity renders \ac{LIV} analyses on pulsars comparatively less sensitive, in particular in the linear scenario. On the other hand, in the quadratic scenario, their fast variability gives them a huge advantage, making them competitive sources for the search of \ac{LIV} effect. Nonetheless, contrary to flaring \acp{AGN} or \acp{GRB}, new observations of pulsar such as the Crab one do not depend on luck but can be carefully planned. This continuous accumulation of new data can lead to a predictable increase of statistics and therefore improved sensitivity of the analysis. This is particularly interesting as the current \ac{ML} analyses are still mainly limited by background fluctuations and systematics such as the pulse shape and its energy evolution in the case of the Crab pulsar. In~\cite{Ahnen:2017wec} the authors stated that a total dataset of $\sim$$2000$\,h on the Crab pulsar is within reach for the \ac{MAGIC} collaboration alone given the regular observations performed for calibration purposes. And this data set could be further enlarged by taking into account the data accumulated on the Crab pulsar by \ac{H.E.S.S.} and \ac{VERITAS} (see also Section~\ref{Subsec:CombiningIACTs}) with the potential of addressing some of the main limitations of the current analyses mentioned above and thus exploring \ac{QG} scales beyond the current best limits, in particular in the quadratic scenario. Lastly, only pulsars with periods of a few tens of milliseconds have been detected with \acp{IACT} so far. Detection of \ac{VHE} gamma rays from pulsars with periods down to a few milliseconds would allow to constrain the emission time more strongly. This would lead to an improvement of the current limits on $E_{\mathrm{QG}}$ by an order of magnitude, further exploiting the potential of these sources in the search for \ac{LIV}.

\section{\label{Sec:Future}An Eye on the Future}

In previous sections, we discussed the evolution of the search for \ac{LIV} effects with \acp{IACT}. The studies, which we reported on, have set strong constraints on the \ac{LIV} energy scale, and significantly restricted the parameter space. More importantly, in those works, diverse ideas were proposed, various analysis methods had been developed, and different effects investigated. 
Ever since the first \ac{LIV} study with an \ac{IACT}, numerous ameliorations have been brought to the field in the form of technical or analysis improvements and the future ahead of us is certainly no exception. 
The most important question at this point is where do we go from here. What can we do to accelerate the research and contribute to the understanding of \ac{QG}?
In this section we try to outline some ideas on what could be the next steps in that regard. 
Hopefully, this will motivate the reader to perform some of the proposed research. We certainly intend to take our part in this endeavour.

\subsection{\label{Subsec:AnalysisRefinement}Refinement of the Analysis Technique}

In Section~\ref{Subsec:MaximumLikelihood} we presented the state of the art \ac{ML} method for testing energy-dependent photon group velocity. Here, we will discuss some possibilities for improving the method and increasing the analysis sensitivity. 

Firstly, let us return to the definition of the likelihood function (Equation~(\ref{eq:Likelihood})) and in particular how individual events are selected and weighted. 
Note that the probability for each event to be a part of the signal, as defined in Equation~(\ref{eq:gammaProb}), is the same for every event. The same is true for the probability for being a part of the background. Moreover, in order to reduce the background, the \ac{IACT} data analysis involves applying cuts on the parameters describing events properties.
This approach inevitably leads to cutting out some of the signal events as well. As we have seen, most of the \ac{LIV} analysis methods rely on individual events. Therefore, cutting out signal events reduces the analysis sensitivity. However, in a recent paper, D'Amico et al. \cite{DAmico:2021qub} proposed an alternative \ac{IACT} data analysis method, which considers all events in the ON region without applying any cut. Instead, \acp{PDF} for the signal and the background are calculated based on the parameters of individual events. This method could, therefore, be used to calculate $p^{(\mathrm{s})}_{i}$ and $p^{(\mathrm{b})}_{i}$ as \acp{PDF}, without discarding any events, whether of signal or background origin.

Secondly, the real strength of the \ac{ML} method relies in its modularity, meaning that the components of the likelihood function listed in Section~\ref{Subsec:MaximumLikelihood} can be refined, and additional terms describing nuisance parameters can be added without limitations. Furthermore, likelihoods from individual targets can easily be combined in a joint likelihood (see \mbox{Section~\ref{Subsec:CombiningIACTs}}).
For example, in case the intrinsic spectrum changes with time, the function $\Phi_{\mathrm{int}}(E)$ can be generalised to $\Phi_{\mathrm{int}}(E, t)$. Additionally, by taking the product $F(t + \eta_{n}E^n)\Phi_{\mathrm{int}}(E)$ in Equation~(\ref{eq:PDF}) we assumed that the emission time $t$ of individual photons does not depend on their energy. 
Indeed, in the \ac{LIV} studies performed so far, there was no strong evidence of changes of spectral shape during \ac{AGN} flaring episodes nor in \acp{GRB} in the \ac{VHE} gamma-ray range. 
Nevertheless, waving this simplification might increase the sensitivity of the \ac{ML} analyses. 
Present day state of the art models used to describe emission from astrophysical sources are nowhere near refined and accurate enough to predict the exact emission time of each particular photon. Hopefully, future emission models will be precise enough to allow creating emission light curve templates simultaneously depending on emission time and energy. E.g., instead of having independent temporal and spectral distributions as $F(t + \eta_{n}E^n) \, \Phi_{\mathrm{int}}(E)$, we could take a two-dimensional distribution as $F_{\mathrm{2D}}(t + \eta_n E^n, E)$ to account for potential source intrinsic delays. Therefore, further progress in the modelling of the sources emission mechanism such as initiated in~\cite{Perennes:2019sjx} will certainly play an important role in future \ac{LIV} studies.

\subsection{\label{Subsec:CombiningIACTs}Combining Data from Different Sources and Instruments}

We have already seen on the example of the gamma-ray absorption (see {Sections \ref{Subsubsec:Biteau_Williams}--\ref{Subsubsec:Lang_et_al}}) how a combination of several sources in a single study improved the sensitivity of the analyses. An equivalent approach is still to be fully applied when testing the photon group velocity. Indeed, a combination of sources observed at a wide range of redshifts is the key to disentangling of intrinsic source effects from a real \ac{LIV} effect. A source-intrinsic energy-dependent photon emission time can mimic an effect of \ac{LIV}, leading to a misinterpretation as energy-dependent time of flight. Alternatively, intrinsic effects could have the same magnitude in the opposite direction from the \ac{LIV} effect, canceling it and preventing a detection; a scenario known as a conspiracy of nature. Nonetheless, a \ac{LIV} effect, if it exists, should be present in all observational data, and directly depend on the distance of a source. A combination of sources at different redshifts would mitigate the potential effect of source intrinsic effects. Furthermore, emission from different types of sources is a result of different physical processes, and subject to different emission dynamics. Therefore, combining different types of sources could further limit the contribution of source-intrinsic effects. 
These two factors, combining data from different types of sources and from observations at different redshifts, are instrumental in the development of a significantly more robust \ac{LIV} analysis.

The likelihood function, can be relatively easily extended to consider additional data sets from the same or other sources. Note that a joint likelihood can be constructed from the product of individual likelihoods, each of them described by Equation~(\ref{eq:Likelihood}). This single analysis of multiple data sets (sharing the same \ac{LIV} parameter $\eta_n$) is possible as long as each data set is accompanied by its set of functions (light curve template, spectral distribution, acceptance, energy resolution, etc.) describing its particularity and its condition of observations. 

Recently, an inter-experiment collaboration has been formed. The so-called \ac{LIV} Consortium assembles researchers from all three currently operating \acp{IACT} experiments (\ac{MAGIC}, \ac{H.E.S.S.}, \ac{VERITAS}) with the goal of combining data from different sources. 
In addition, the \ac{LIV} Consortium is working on unifying observational data from different \acp{IACT} facilities. This will immediately give access to a notably larger pool of sources than what has been used so far in \ac{LIV} studies with \acp{IACT}. 
Furthermore, combining different instruments in a single analysis, with a particular consideration of individual instrument response functions, is expected to decrease systematic uncertainties.
Finally, such combination effort provides the necessary environment to harmonize the details of the analysis. As we have previously seen, analysis techniques can slightly differ from one experiment to the other. Combining individual best practices will further contribute to the research efficiency.
Therefore, a combination of observations from all three currently operating \acp{IACT} will provide a major improvement in the constraints on the \ac{LIV} energy scale, not specifically on the facial value of the limits, but more importantly on the robustness of the results. The work done by the \ac{LIV} Consortium can also be regarded as preparatory activities for the \ac{CTA} era, which we discuss in the following paragraph. 
Preliminary results have been presented in conferences \cite{2017ICRC...35..646N}, while the final results are expected soon.

Another improvement in the direction of combining instruments would be to extend data samples to lower and higher energies, e.g., \textit{Fermi}-\ac{LAT} for the MeV--GeV energy range and \ac{HAWC} or \ac{LHAASO} for energies above PeV. These experiments provide complementary information to the one of \acp{IACT}. It thus makes sense to combine the observations from these experiments to further increase the sensitivity of tests of \ac{LIV} on gamma rays. 
Possible benefits are quite tantalizing, although such endeavor would be far from trivial, especially when it comes to treatment of different instrumental effects. 

However, near future holds the prospect of the \ac{CTA}, which will be an order of magnitude more sensitive than any of the existing Cherenkov telescopes \cite{CTAperformance}. 
Combined with a large number of telescopes, located in both hemispheres, the \ac{CTA} will cover larger portions of the sky and observe more sources. 
This is particularly noteworthy for transient events, such as \acp{GRB} and \ac{AGN} flares, and less bright sources, such as pulsars, all of which are essential for \ac{LIV} studies. So far, only one \ac{LIV} study was performed on a \ac{GRB} observed with an \ac{IACT}, and only a total of four \acp{GRB} have been observed with \acp{IACT} until now. The \ac{CTA} is expected to improve on this statistics. 
In Section~\ref{Subsec:Fermi-LAT}, we discussed how Vasileiou et al. considered four \acp{GRB} in their study. GRB~090510, although located at the lowest redshift, yielded constraints on $E_\mathrm{QG}$ stronger by a factor of $2\text{--}20$ in the linear and $3\text{--}15$ in the quadratic scenarios, compared to results from other three \acp{GRB}  \cite{Vasileiou:2013vra}. This was achieved due to a combination of the highest energy in the sample and the fastest variability. 
In Section~\ref{Sec:ResultsComparison} we compared the results obtained on GRB~090510 to the ones from GRB~190114C. The first one was a short \ac{GRB}, with more than double the redshift of the second one. The signal from GRB~190114C was detected at two orders of magnitude higher energies, however the MAGIC telescopes detected mostly the afterglow phase. With the \ac{CTA}, we hope to detect prompt emission as well, which is expected to be a more variable phase of \acp{GRB}. 
Furthermore, the \ac{CTA} will extend the range of accessible energies both to lower (down to 20\,GeV) and higher bands (up to 300\,TeV). 
The highest energies detectable with the \ac{CTA} will be almost an order of magnitude higher than the the ones accessible with the currently operating \acp{IACT}. 
On the other hand, extending the observation window to lower energies, will grant access to gamma rays not absorbed by the \ac{EBL}, enabling observations of sources at higher redshifts. Referring again to Table~\ref{Tab:TOFsensitivity}, higher redshift of a source improves the sensitivity to the \ac{LIV} energy scale proportionally to the redshift in the linear, and somewhat less in the quadratic contribution. Alas, these two improvements cannot be combined. Gamma rays with the highest energies, emitted from the highest redshifts, will be absorbed by \ac{EBL} before reaching Earth, so only one of these advantages will be accessible at a time. Nevertheless, there is an improvement coming from the wide energy range itself. Whichever effect of \ac{LIV} is tested for (time of flight, universe transparency, etc.), the flux at the highest energies is compared to what is assumed to be the intrinsic emission. The latter is estimated from observations in the lower energy bands. 
\ac{LIV}, if real, affects the gamma rays in the lower energy band as well the most energetic events. Granted, the effect is smaller for low energy band, but still present. Assuming that low energy photons are unaffected by the \ac{LIV} induces a bias, and, ultimately, decreases the sensitivity of analyses. The bias will be smaller for a wider gap between the two energy bands used. 
Observations in the lower energy band can be obtained either using the same instrument, as in the case of e.g., PKS~2155-304 (see \cite{2011APh....34..738H} and Section~\ref{Subsec:PKS2155_HESS}), or from other instruments combined with theoretical inferences, as in the case of GRB~190114C (see \cite{Acciari:2020kpi} and Section~\ref{Subsec:GRB190114C}). Ideally, observations in both energy bands would be performed with the same instrument with a wide range of observable energies. The former would reduce possible systematic effects, while the latter would decrease the potential \ac{LIV}-induced bias. The \ac{CTA}, with the range of accessible energies spanning over more than four orders of magnitude, will answer this need. 
While it is difficult to predict the light curves and spectra that the \ac{CTA} will observe from \acp{GRB}, we can draw some conclusions by extrapolating the case of GRB~090510 to higher energies. Assuming that gamma rays emitted during the \acp{GRB} prompt phases can reach energies as high as few hundred GeV to a few TeV, the sensitivity to \ac{LIV} effects would increase by one to two orders of magnitude. Even if we have to settle with lower redshifts, thus somewhat reducing the sensitivity (see Table~\ref{Tab:TOFsensitivity}; e.g., for $z=0.3$ the loss of sensitivity is at most a factor of $\sim$$3$ compared to $z=0.9$, the redshift of GRB~090510), the gain would still be substantial. 
A similar reasoning can be applied to studies based on spectral analysis. By lowering the detection energy threshold, the intrinsic spectra will be more precisely determined at the lower energy end of spectra, leading to better spectral fitting, and thus decreasing the uncertainties on the \ac{LIV} energy scale. 
The \ac{CTA} Consortium has published a projection of the \ac{CTA}’s capabilities for probing fundamental physics, including \ac{LIV}~\cite{CTA:2020hii}. The study was limited to universe transparency to gamma rays. The authors estimated that the \ac{CTA} will be able to probe the \ac{LIV} energy scale a factor of two to three higher than the most sensitive studies published so far, whether based on a single source, or combining several sources.

\subsection{\label{Subsec:OtherLIVeffects}Additional and Alternative Lorentz Invariance Violation Effects and Related Phenomena}

Apart from the numerous interesting studies described in the previous sections, there are still a plethora of \ac{LIV}-induced effects and related phenomena which have not been studied with \acp{IACT} data. We will briefly mention those here. 

Firstly, the energy-dependent arrival time delay between two simultaneously emitted photons from the same source is calculated from Equations~(\ref{eq:TimeDelay1}) and (\ref{eq:LIVComovingDistance}). These were derived taking a comoving trajectories of the photons and their respective energy-dependent velocities. As mentioned in Section~\ref{Sec:TOF}, one could start the derivation from a modified general relativistic dispersion relation \cite{Pfeifer:2018pty}, or by modifying spacetime translations with the photon dispersion relation \cite{Rosati:2015pga}, and obtain different results for the photon time of flight. It would be interesting to investigate the differences between these models on \acp{IACT} data.

Secondly, all the tests of energy-dependent photon group velocity performed on \acp{IACT} data were based on Equation~(\ref{eq:photonvelocity}), which is deterministic in the sense that it assumes that all photons of the same energy will propagate with the same group velocity. However, there are models which propose fluctuations of the photon group velocity as a consequence of fluctuations of the spacetime foam \cite{Ellis:1999uh}. This phenomenon, often referred to as stochastic \ac{LIV} models photon group velocity as \cite{Vasileiou:2015wja}:

\begin{equation}\label{eq:StochasticLIV}
    v_\gamma (E) = c + \delta v_\gamma (E),
\end{equation}
where the velocity modification $\delta v_\gamma (E)$ is randomly distributed according to normal distribution with mean in zero, and a standard deviation given with:
\vspace{6pt}
\begin{equation}\label{eq:StochasticLIVsigma}
    \sigma_n (E) = c \frac{1+n}{2} \left( \frac{E}{E_{\mathrm{QG,}n}} \right)^n.
\end{equation}

The stochastic \ac{LIV} was tested on \textit{Fermi}-\ac{LAT} observation of GRB~090510 \cite{Vasileiou:2015wja}. Only linear term was constrained to $E_{\mathrm{QG,1}} > 3.4 \times 10^{19}$\,GeV. So far, no similar study was performed on \acp{IACT} observations. As pointed out by Bolmont in \cite{bolmont:tel-01388037}, distant pulsars with sharp pulsation peaks would be excellent probes of stochastic \ac{LIV}.

A substantial portion of this work was dedicated to effects \ac{LIV} has on photon interactions. An important process in astrophysical sources of gamma rays is (inverse) Compton scattering. {Abdalla \& B\"{o}ttcher} analysed a possible influence of \ac{LIV} on Compton scattering in \cite{Abdalla:2018sxi} and concluded that \ac{LIV} signatures were expected to be important only for incoming gamma-ray energies above $\sim$$1$\,PeV. While in light of the recent results from \ac{LHAASO} (see Section~\ref{Subsec:PhotonDecay}) this effect might draw some attention, {Abdalla \& B\"{o}ttcher} also concluded that even in the superluminal \ac{LIV} scenario, the Klein--Nishina cross section would still be strongly dominated by the Thomson cross section, while in the subluminal scenario, the Klein--Nishina cross section would be even more strongly suppressed. Overall conclusion was that an \ac{LIV} modified Compton scattering was not likely to be relevant in realistic astrophysical situations.

We would also like to point out that all investigations of universe transparency for gamma rays were performed on photons of energies up to 100\,TeV. Remembering again the \ac{LHAASO} detection of a photon of $\gtrsim 1$\,PeV, we are strongly encouraged to extend our test to higher gamma-ray energies. As we have demonstrated in Section~\ref{Subsec:UniverseTransparency}, and in particular in Figure~\ref{Fig:Energy_Threshold_SED}, this will extend the photon target field to the \ac{CMB}, which is substantially more dense than any of the \ac{EBL} components. Moreover, in Figure~\ref{Fig:crossection2}, we have shown how the shape of the Breit--Wheeler cross section deforms in \ac{LIV} subluminal scenario. These deformations were not significant for gamma-rays up to $\sim$$100$\,TeV, but might play an important role for higher gamma-ray energies.

As previously mentioned, a study done by the \ac{H.E.S.S.} Collaboration \cite{Abdalla:2019krx} is the only one in which time of flight and universe transparency tests were performed on the same data set (see Sections \ref{Subsec:Mrk501_HESS} and \ref{Subsubsec:HESS_spectral}). However, these two effects were tested independently of each other. In fact, there has never been a study that combined two different \ac{LIV} effects.
We have argued in Section~\ref{Sec:ResultsComparison} why strong limits based on one \ac{LIV} effect do not necessarily constrain other effects. For example, limits on $E_{\mathrm{QG}}$ based on the universe transparency do not apply to the photon group velocity. Excluding one effect does not automatically exclude all \ac{LIV} effects. Nonetheless, considering \ac{VHE} gamma rays from astrophysical sources, it seems natural to wonder what would be the net observational result if several \ac{LIV} effects were present.
Namely, in Equation~(\ref{eq:PDF}), term $F(E)$ takes into account propagation effects, such as \ac{EBL} absorption, but under the assumption that photon interactions were not \ac{LIV} affected. A combined effect study would presume that both photon group velocity and photon interactions were affected by the \ac{LIV}. Different effects could manifest on the same, or different energy scales. Since there is no fully formulated theory of \ac{QG}, and we do not know what the effects of \ac{QG} are, this test is well justified. Implementation of such a test could be realised by modifying terms in the signal \ac{PDF} in Equation~(\ref{eq:PDF}). This might present a significant challenge, especially a computational one when maximising the likelihood function, and in particular if an independent value of $E_{\mathrm{QG}}$ is associated to each considered \ac{LIV} effect.

Finally, there are possible effects of \ac{LIV}, which have not been mentioned earlier, such as vacuum birefringence, vacuum dispersion, and vacuum anisotropy \cite{Kostelecky:2008be}. 
The vacuum birefringence implies that the polarization vector of a linearly polarized photon will rotate dependently on the energy of the photon. As a consequence, a linearly polarised signal from an astrophysical source will be depolarised by the time it reaches Earth. Therefore, measurements of polarisation degree in a signal would constrain the vacuum birefringence effect (see, e.g., refs. \cite{Kislat:2017kbh,Friedman:2020bxa} for studies performed on optical observations, and refs. \cite{Toma:2012xa,Kostelecky:2013rv,Gotz:2014vza} for studies performed on hard X-ray -- soft gamma-ray observations.). These results are substantially more constraining than the tests based on modified photon kinematics. 
The ratio of sensitivities of the birefringence tests and the time of flight tests, performed on the same data sample, is proportional to the energy of the photons in the sample, as discussed by {Kosteleck\'{y} \& Mewes} in \cite{Kostelecky:2009zp}. As they put it descriptively, in order to achieve sensitivity in the time of flight tests comparable to the birefringence tests, one would need to achieve the timing resolution on the order of the inverse frequency of the photons. Even if this was instrumentally feasible (which is not), the measurement precision would be spoiled by the inability to constrain the photon emission times. 
Unfortunately, the \ac{IACT} detection technique does not allow for measurements of gamma-ray polarisation. Recently, a novel satellite-based gamma-ray detector was proposed, which would be capable of measuring polarisation \cite{DeAngelis:2016slk} of gamma rays in a lower energy band. Unfortunately, the proposal was not accepted, but there is hope it will be selected for some future space mission, or that a similar concept such as AMEGO\footnote{\textls[-15]{All-sky Medium Energy Gamma-ray Observatory (AMEGO, \href{https://asd.gsfc.nasa.gov/amego/index.html}{https://asd.gsfc.nasa.gov/amego/index.html}, accessed 15 July 2021) is a concept for the MeV sky~exploration.}} will be realised. X-ray polarimetry will certainly gain with soon to be launched IXPE\footnote{Imaging X-ray Polarimetry Explorer (IXPE, \href{https://ixpe.msfc.nasa.gov/index.html}{https://ixpe.msfc.nasa.gov/index.html}, accessed 15 July 2021).}. 

Given that \acp{IACT} cannot measure the polarisation of gamma rays, a broad discussion of birefringence would be out of the scope of this work. Nevertheless, considering strong constraints on \ac{LIV} based on birefringence tests, we should note that these do not render time of flight tests useless. Indeed, already from theoretical considerations it is clear that in some cases the energy-dependent photon group velocity does not imply vacuum birefringence \cite{Kostelecky:2009zp} (see also \cite{bolmont:tel-01388037} for a critical discussion). Therefore, strong limits on the vacuum birefringence do not necessarily constrain the energy dependence of the photon group velocity. 
Furthermore, while \acp{IACT} are not convenient for tests on photon polarisation, one should keep in mind that, as remarked in \cite{2013ICRC...33.2768Z}, vacuum anisotropy could be constrained by performing \ac{LIV} tests on sources in different directions. Hence, we should not be satisfied with setting stronger and stronger constraints on the \ac{LIV} energy scale only. We should also strive towards building a rich statistics of sources located at different directions, as well as at different redshifts, as we already argued. Again, the \ac{CTA} is expected to considerably contribute in this respect.

\section{Conclusions}
\label{Sec:Conclusions}
Most of investigators agree that there is a fundamental, quantum description of gravity. Though not formulated yet, the theory of \ac{QG} is expected to resolve what happens in extreme gravitational potentials, such as singularities within black holes predicted by the general theory of relativity, or early universe, but also to push us in the direction of formulating the next unification theory describing all interactions. 
The expected realm of \ac{QG} is the Planck scale, far above the reach of any physical laboratories present today, or any experiment envisaged in the near future. Nevertheless, some investigators believe even \ac{VHE} gamma rays from astrophysical sources would feel minuscule effects of \ac{QG}. 
These effects, consequences of the so-called \ac{LIV}, would manifest as the photon group velocity deviating from the Lorentz invariant speed of light $c$, or as modified photon interactions. 
Given the cosmological distances gamma rays cover to reach Earth, there is a hope that these effects would accumulate sufficiently enough to be detected with gamma-ray detectors.

In this review we presented all experimental tests of \ac{LIV} performed with \acp{IACT}. We followed the historical development of the field. However, in order not to create a confusion, we first covered tests of energy-dependent photon group velocity, and then tests of modified photon interactions. A strictly chronological overview of the results was given in Table~\ref{Tab:Results}. 
24 years have past since the first proposal that gamma rays emitted from astrophysical objects could be used to search for effects of \ac{QG}. In fact, the proposal singled out \acp{GRB} as gamma-ray sources to be used for these tests. Only two years later, the first study using \ac{IACT} Whipple was published on data from a blazar Mrk~421. Meanwhile, twenty more years had to pass before we were able to test the \ac{LIV} on GRB~190114C observed by \ac{MAGIC}. 
So far, no significant violation of the Lorentz symmetry was detected. However, in the past 22 years since the first experimental result was published, quite strict bounds were set on the level of \ac{LIV}. In particular, considering the linear term in modified photon dispersion relation (Equation~(\ref{eq:moddispastro})), the lower limit on $E_{\mathrm{QG,1}}$ has surpassed the Planck energy. This is especially true when photon interactions are considered. While the expected energy level of \ac{QG} is indeed the Planck scale, there is no strictly defined interval for $E_{\mathrm{QG}}$, so the test will continue. 
When it comes to the quadratic term in Equation~(\ref{eq:moddispastro}), there is still quite some parameter space of $E_{\mathrm{QG,2}}$ to be investigated, both considering the photon group velocity and photon interactions; and in Section~\ref{Sec:ResultsComparison} we discussed why it is important to test for all possible effects, as well as both for subluminal and superluminal scenarios. 

We hope to have demonstrated that \acp{IACT} played an important role in search for \ac{LIV} and setting strong constraints on its energy scale. More importantly, we tried to argue that \acp{IACT} still have much to say in this field, and that observations with these instruments will ultimately lead to either detecting \ac{LIV}, or confirming that the universe remains Lorentz invariant up to trans-Planckian energies.
In Section~\ref{Sec:Future} we presented our vision of the future development of this field. It goes without saying that we have great expectations of future facilities. In particular, the \ac{CTA}, which promises a detection of several \acp{GRB} each year. However, new instruments will not do on their own, and additional improvements and development of analysis techniques will be necessary. In addition, there are hypothesised phenomena (e.g., stochastic \ac{LIV}) which have not been tested for yet. Obviously, there is quite some work cut out for us. Whether we detect some effect of \ac{LIV}, or it turns out that the Lorentz symmetry is perfectly preserved, one thing is for sure: exciting times are ahead.

\vspace{6pt} 
\authorcontributions{All authors contributed equally to the writing. All authors have read and agreed to the published version of the manuscript.}
\funding{T.T. and J.S. acknowledge funding from the University of Rijeka, project number 13.12.1.3.02. T.T. also acknowledges funding from the Croatian Science Foundation (HrZZ), project number IP-2016-06-9782.
D.K. acknowledges funding from the European Union's Horizon 2020 research and innovation programme under the Marie Sk\l{}odowska-Curie grant agreement No. 754510, from the ERDF under the Spanish Ministerio de Ciencia e Innovaci\'{o}n (MICINN), grant PID2019-107847RB-C41, and from the CERCA program of the Generalitat de~Catalunya.}

\acknowledgments{We would like to thank G. Bonnoli, C. Levy, M. Mart\'{i}nez, H. Mart\'{i}nez-Huerta, D. Paneque, and F. Tavecchio for fruitful discussions, and verification of historical facts. The authors acknowledge networking support by the COST Action CA18108.}

\conflictsofinterest{The authors declare no conflict of interest.}

\section*{\hypertarget{sec:Abbreviations}{Abbreviations}}\label{Sec:Abbreviations}
\noindent{The following acronyms and abbreviations are used in this manuscript:
\begingroup
\begin{acronym}
    \acro{AGN}{active galactic nucleus}
    \acro{B-H}{Bethe--Heitler}
    \acro{CMB}{cosmic microwave background}
    \acro{CTA}{Cherenkov Telescope Array}
    \acro{C.U.}{Crab units}
    \acro{CWT}{continuous wavelet transform}
    \acro{DisCan}{dispersion cancellation}
    \acro{DSR}{Doubly Special Relativity}
    \acro{EBL}{extragalactic background light}
    \acro{ECF}{energy cost function}
    \acro{FACT}{First G-APD Cherenkov Telescope}
    \acro{GRB}{gamma-ray burst}
    \acro{GBM}{Gamma-ray Burst Monitor}
    \acro{HAWC}{High Altitude Water Cherenkov}
    \acro{HEGRA}{High Energy Gamma Ray Astronomy}
    \acro{H.E.S.S.}{High Energy Stereoscopic System}
    \acro{IACT}{imaging atmospheric Cherenkov telescope}
    \acro{LAT}{Large Area Telescope}
    \acro{LHAASO}{Large High Altitude Air Shower Observatory}
    \acro{LIV}{Lorentz invariance violation}
    \acro{MAGIC}{Major Atmospheric Gamma Imaging Cherenkov}
    \acro{MCCF}{modified cross correlation function}
    \acro{MD}{minimal dispersion}
    \acro{ML}{maximum likelihood}
    \acro{PC}{peak comparison}
    \acro{PDF}{probability distribution function}
    \acro{PV}{PairView}
    \acro{RB}{radio background}
    \acro{QG}{Quantum Gravity}
    \acro{SMM}{sharpness maximisation method}
    \acro{TS}{test statistic}
    \acro{VERITAS}{Very Energetic Radiation Imaging Telescope Array System}
    \acro{VHE}{very high energy \acroextra{(100\,GeV\,$<E<$\,100\,TeV)}}
\end{acronym}
\newacroplural{IACT}{imaging atmospheric Cherenkov telescopes}
\newacroplural{AGN}[AGN]{active galactic nuclei}
\endgroup
}

\end{paracol}
\reftitle{\hypertarget{sec:References}{References}}\label{Sec:References}
\externalbibliography{yes}
\bibliography{main}

\end{document}